\documentclass[a4paper, 12pt]{article}

\usepackage{AcdemicWriting}
\usepackage{mathtools}

\nolinenumbers
\pdfoutput = 1 
\date{}

\title{Modeling Multivariate Outcomes with Dependence Structures of Interconnected Modules: Evaluating the Effect of Alcohol Intake on Plasma Metabolomics}
\author{
	Yifan Yang \thanks{Address: Department of Population and Quantitative Health Sciences, Case Western Reserve University, Cleveland, Ohio, 44106 U.S.A., 
		Email: yiorfun@case.edu}
	\ \ \ \
	Chixiang Chen \thanks{Address: School of Medicine, University of Maryland, Baltimore, Maryland, 21201 U.S.A., 
		Email: chixiang.chen@som.umaryland.edu}
	\ \ \ \
	Hwiyoung Lee \thanks{Address: School of Medicine, University of Maryland, Baltimore, Maryland, 21201 U.S.A., 
		Email: Hwiyoung.Lee@som.umaryland.edu}
	\ \ \ \
	Ming Wang \thanks{Address: Department of Population and Quantitative Health Sciences, Case Western Reserve University, Cleveland, Ohio, 44106 U.S.A., 
		Email: mxw827@case.edu}
	\ \ \ \
	Shuo Chen \thanks{Address: School of Medicine, University of Maryland, Baltimore, Maryland, 21201 U.S.A., 
		Email: shuochen@som.umaryland.edu}
}

\date{}

\begin{document}
	
	\maketitle

	\begin{abstract}	
		Alcohol consumption has been shown to influence cardiovascular mechanisms in humans, leading to observable alterations in the plasma metabolomic profile. 
		Regression models are commonly employed to investigate these effects, treating metabolomics features as the outcomes and alcohol intake as the exposure. 
		Given the latent dependence structure among the numerous metabolomic features (e.g., co-expression networks with interconnected modules), addressing this structure is crucial for accurately identifying metabolomic features associated with alcohol intake.
		However, integrating dependence structures into regression models remains difficult in both estimation and inference procedures due to their large or high dimensionality.
		To bridge this gap, we propose an innovative multivariate regression model that accounts for correlations among outcome features by incorporating an interconnected community structure. 
		Furthermore, we derive closed-form and likelihood-based estimators, accompanied by explicit exact and explicit asymptotic covariance matrix estimators, respectively.
		Simulation analysis demonstrates that our approach provides accurate estimation of both dependence and regression coefficients, and enhances sensitivity while maintaining a well-controlled discovery rate, as evidenced through benchmarking against existing regression models. 
		Finally, we apply our approach to assess the impact of alcohol intake on $249$ metabolomic biomarkers measured using nuclear magnetic resonance spectroscopy. 
		The results indicate that alcohol intake can elevate high-density lipoprotein levels by enhancing the transport rate of Apolipoproteins A1.
		
		\smallskip
		
	\end{abstract}
	
	\clearpage
	
	\pagenumbering{arabic}

	\section{Introduction}
	
	\label{Sec:introduction}
	
	Simultaneous measurement of hundreds of thousands of biological features has revealed complex scientific mechanisms across various fields \citep{HeKangHong2019, KeRenQi2022}. 
	Numerous statistical methodologies have been developed to address the challenges associated with high-dimensional data analysis. 
	For instance, shrinkage or penalty techniques have been employed in linear regression models, leading to the establishment of theoretical properties for the sparse estimators \citep{EfronHastieJohnstone2004, HastieTibshiraniWainwright2015}. 
	Analogous regularization methods have successfully extended to estimate large-scale covariance or precision matrices, incorporating the assumption of bandability, sparsity, or low-rank \citep{WuPourahmadi2003, BickelLevina2008a, BickelLevina2008b, FanLiaoMincheva2013}.
	From an inference perspective, multiplicity-adjusted procedures have been proposed to enable the simultaneous testing of numerous hypotheses, irrespective of whether the underlying test statistics are independent \citep{StoreyTaylorSiegmund2004, Efron2004} or heavily correlated \citep{LeekStorey2008, FanHanGu2012, FanHan2017}.
	
	Conventionally, statistical methods handling high-dimensional variables can be broadly categorized into two classes. 
	In the first class, high-dimensional variables are considered as predictors.
	Commonly, regression shrinkage methods, e.g., the lasso \citep{Tibshirani1996} and its many variants \citep{YuanLin2007, FanLv2008, HeKangHong2019}, are utilized to select the variables. 
	The second class focuses on high-dimensional variables as multivariate outcomes of regression. 
	For instance, high-dimensional imaging data are often modeled as outcomes while considering spatial dependence. 
	Additionally, multiple testing methods widely employed in omics data analysis fall into this category \citep{LeekStorey2008, FanHanGu2012, FanHan2017}.
	In this paper, our emphasis is on the second category of data analysis.
	We are motivated to assess the effect of alcohol intake on metabolomic profiles.
	Specifically, our goal is to incorporate the structured dependence among the features in multivariate outcomes into a regression model linked to exposures.
	
\begin{figure}[!h]
		\centering
		\includegraphics[width = 1\linewidth]{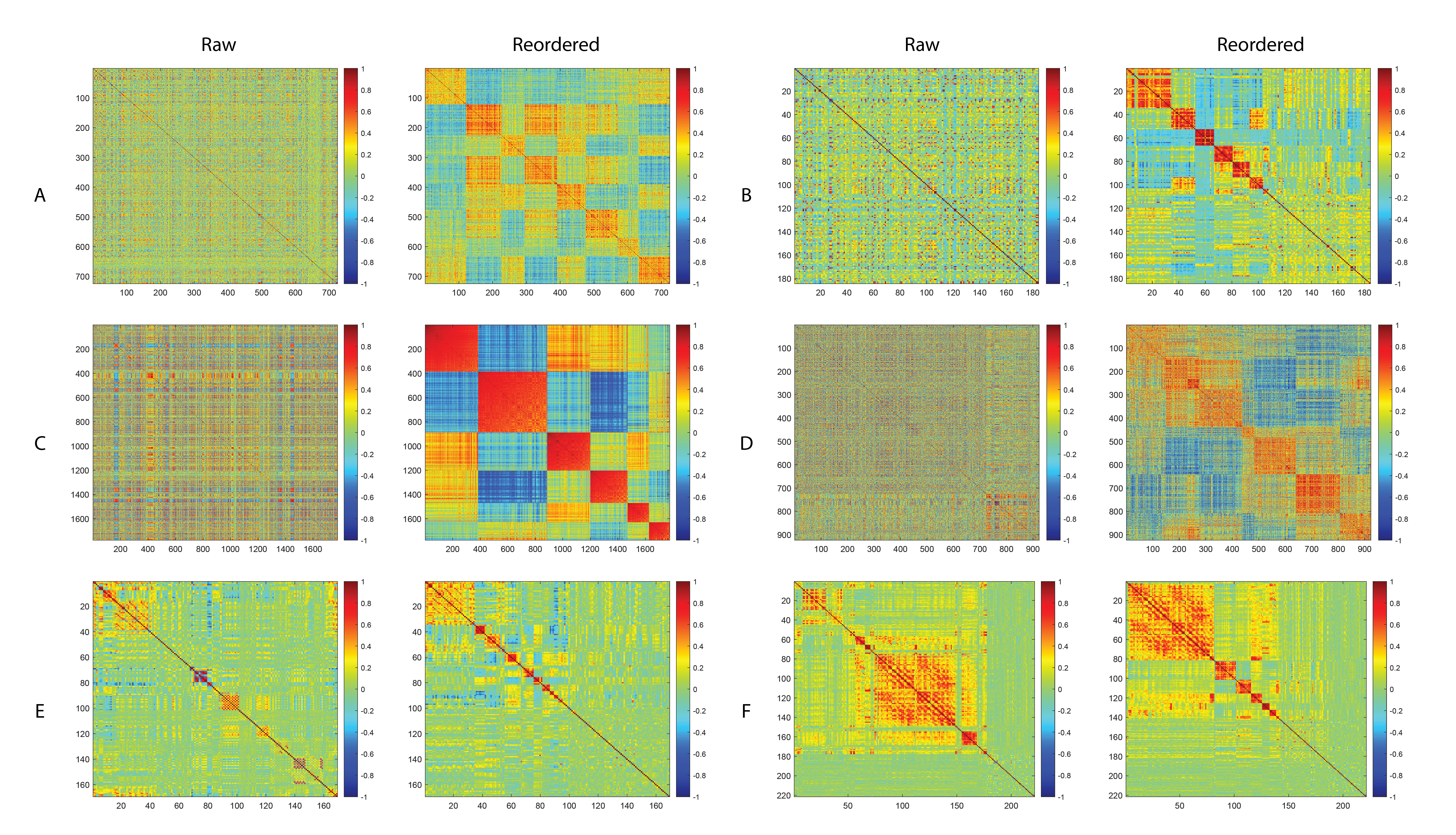} 
		\caption{We illustrate interconnected community structures with or without singletons across various datasets. 
			A: the heat maps of raw sample correlation matrix (left) and reordered sample correlation matrix (right) based on the genomics dataset \citep{SpellmanSherlockZhang1998}; 
			B: the heat maps of raw sample correlation matrix (left) and reordered sample correlation matrix (right) based on the proteomics dataset \citep{YildizShyrRahman2007};
			C: the heat maps of raw sample correlation matrix (left) and reordered sample correlation matrix (right) based on the gene expression dataset \citep{RitchieSurendranKarthikeyan2023}; 
			D: the heat maps of raw sample correlation matrix (left) and reordered sample correlation matrix (right) based on the multi-omics dataset \citep{PerrotLevyRajjou2022};
			E: the heat maps of raw sample correlation matrix (left) and reordered sample correlation matrix (right) based on the environmental exposome dataset \citep{ISG2021};
			F: the heat maps of raw sample correlation matrix (left) and reordered sample correlation matrix (right) based on the environmental plasma metabolomics dataset \citep{ISG2021}.
		}
		\label{Fig:ub_examples}
	\end{figure}

\begin{figure}[h]
		\centering
		\includegraphics[width = 0.8\linewidth]{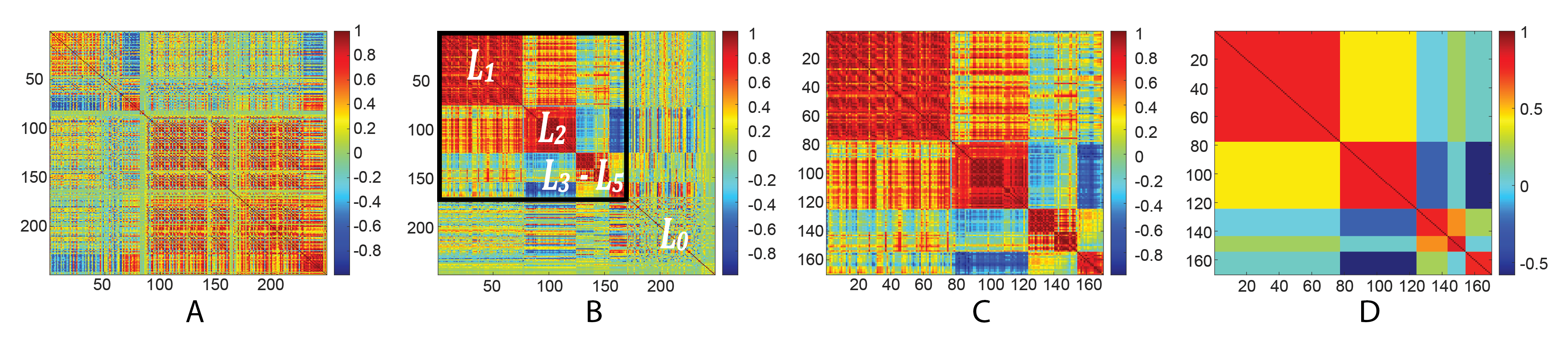} 
		\caption{We illustrate the interconnected community structure with singletons in NMR data. 
			A: the heat map of the $249$ by $249$ sample correlation matrix of residuals derived from the raw NMR dataset \citep{RitchieSurendranKarthikeyan2023};
			B: the heat map of the $249$ by $249$ sample correlation matrix of residuals, reordered using a community detection algorithm \citep{ChenZhangWu2024}, exhibiting an interconnected community structure (with singletons);
			C: the heat map of the $170$ by $170$ sample correlation matrix of residuals extracted from the region outlined by the black frame in B;
			D: the estimated population correlation matrix demonstrating an interconnected community structure with ($5$ by $5$) uniform blocks (without any singletons).
		}
		\label{Fig:NMR}
	\end{figure}

	Incorporating the dependence structure into a multivariate regression model poses a twofold challenge. 
	Firstly, the pattern of dependence structure can be latent and intricate, necessitating its detection or estimation before modeling. 
	For example, this challenge is encountered in gene co-expression network analysis \citep{WuMaLiu2021} and multi-omics analysis \citep{PerrotLevyRajjou2022}. 
	Furthermore, effectively utilizing this detected dependence structure in a multivariate regression model is challenging. This is primarily due to the covariance structure of (marginal) outcomes implicitly depending on the detected pattern.
	In our empirical analyses, spanning diverse high-dimensional datasets, including gene expression, proteomics, brain imaging, multi-omics, and more \citep{YangChenChen2024}, a prevalent structure known as the \emph{interconnected community structure} has been identified, as illustrated in Figure~\ref{Fig:ub_examples}.
	Despite the latent existence of interconnected community structures, recent advancements in network structure detection can reliably and consistently unveil them from data \citep{WangLiangJi2020, WuMaLiu2021, LiLeiBhattacharyya2022}.
	%Specifically, this structure possesses characteristics: it is latent, resulting from the application of network detection algorithms to raw data; it is non-sparse with the elements of the population covariance or correlation matrix having small but non-zero values; it exhibits an almost constant-valued block form, where elements within each block display low variability.
	Additionally, it may include numerous singletons or isolated nodes, as illustrated in Figure~\ref{Fig:NMR} (B).
	However, as discussed, due to the existence of the structured covariance or correlation matrix in the data, it is essential to incorporate the corresponding structured dependence into a statistical model to obtain reliable and accurate inference.
	
	%%%%%%%%%%%%%%%%%%
	%%%% Figure 1 %%%%
	%%%%%%%%%%%%%%%%%%
	%%%%%%%%%%%%%%%%%%%%%
	%%%End of Figure 1 %%
	%%%%%%%%%%%%%%%%%%%%%
	
	%%%%%%%%%%%%%%%%%%
	%%%% Figure 2 %%%%
	%%%%%%%%%%%%%%%%%%
	%%%%%%%%%%%%%%%%%%%%%
	%%%End of Figure 2 %%
	%%%%%%%%%%%%%%%%%%%%%

	To address the above gap, we propose the \emph{Multivariate Autoregressive regression model with Uniform-block Dependence} (MAUD).
	This model incorporates the latent interconnected community structure found in data into a parametric regression model. Moreover, we devise an estimation procedure for both dependence parameters and regression coefficients. Additionally, we establish the finite- and large-sample properties of the proposed estimators.

	The MAUD offers noteworthy contributions in three key aspects. 
	Firstly, biological features in outcomes are categorized into distinct communities (groups or clusters) based on preliminary studies, leveraging network detection algorithms.
	Moreover, the MAUD straightforwardly defines dependence parameters both within and between these communities.
	Secondly, the regression coefficient estimators based on the ordinary least-square (OLS) method, the generalized least-square (GLS) method, and the feasible generalized least-square (FGLS) method, are identical. 
	Simultaneously, the dependence parameter estimator is derived through the maximum likelihood (ML) method.
	The third contribution lies in the closed-form exact covariance matrix for the regression coefficient estimator and the closed-form asymptotic covariance matrix for the dependence parameter estimator.
	These enable Wald-type hypothesis tests about all unknown parameters.
	In summary, compared to the conventional regression models that adopt a diagonal or block-diagonal structure to represent dependence in outcome features (thus ignoring non-null off-diagonal correlations), the proposed MAUD offers enhanced accuracy in estimates and inferences.

	This paper proceeds with Section \ref{Sec:methods}, focusing on our methodology for modeling dependence, estimation, and statistical inference. 
	In Section~\ref{Subsec:SAR}, we introduce a simultaneous autoregressive regression model for multivariate outcomes.
	Following that, Section~\ref{Subsec:maud} details the multivariate autoregressive regression model with an interconnected community structure and establishes the relationship between the dependence parameters and the covariance matrix of marginal outcomes.
	Section~\ref{Subsec:estimation} outlines the estimation procedure along with its theoretical properties and Section~\ref{Subsec:inference} discusses hypothesis tests for both the regression coefficients and dependence parameters. 
	In Section \ref{Sec:simulation}, we conduct simulation studies to evaluate the performance of the proposed methodology.
	We then apply the proposed method to real data in Section~\ref{Sec:real} and provide a comprehensive discussion in Section~\ref{Sec:discussion}.
	All proofs, additional definitions, properties, figures, and tables can be found in the \href{Supplementary Material.pdf}{Supplementary Material}.

	\section{Methodology}
	
	\label{Sec:methods}
	
	\subsection{A simultaneous autoregressive regression model}
	
	\label{Subsec:SAR}
	
	Before introducing our primary model (not involving singletons), we begin with a straightforward model motivated by scenarios of interconnected community structures that include singletons.
	Throughout the paper, we use the notation $*$ to emphasize a variable associated with the singletons.
	 
	Let $\textbf{Y} ^ * = \left(\textbf{y}_1 ^ *, \ldots, \textbf{y}_n ^ * \right)$ represent the $S$ by $n$ data matrix, where $\textbf{y}_{i} ^ * = \left(y_{i1}, \ldots, y_{iS} \right) ^ \top$, $i = 1, \ldots, n$, are $n$ independent and identically distributed (i.i.d.) copies of the $S$-dimensional features in outcomes $\textbf{y} ^ * = \left(y_{1}, \ldots, y_{S} \right) ^ \top$.
	We assume that the covariance structure of $\textbf{y} ^ *$ follows an interconnected community structure with singletons, as illustrated in Figure~\ref{Fig:NMR} (B). 
	Specifically, 
	suppose among these $S$ features, there exist $R < S$ features forming $G$ interconnected communities.
	In other words, without loss of generality, the interconnected community structure with singletons uniquely implies a community assigning function $\phi ^ *: \{1, 2, \ldots, S\} \to \{0, 1, \ldots, G\}$: these $R$ features can be assigned to $G$ mutually exclusive and exhaustive communities, i.e., $\phi ^ * (r) = g = 1$ for $r \in \{r_{1, 1}, \ldots, r_{1, L_1}\} \subseteq \{1, \ldots, S\}$, $\phi ^ * (r) = g = 2$ for $r \in \{r_{2, 1}, \ldots, r_{2, L_2}\} \subseteq \{1, \ldots, S\}$, and so forth, $\phi ^ * (r) = g = G$ for $r \in \{r_{G, 1}, \ldots, r_{G, L_G}\} \subseteq \{1, \ldots, S\}$; while the remaining features do not belong to any communities (i.e., singletons), $\phi ^ * (r) = 0$ for $r \in \{1, 2, \ldots, S\} / \cup_{g = 1} ^ {G}\{r_{g, 1}, \ldots, r_{g, L_g}\}$, where we let $L_g > 1$ denote the cardinality of community $g$ for $g = 1, \ldots, G$.
	Moreover, we let $L_0$ denote the number of singletons, such that $S = R + L_0$ with $R = L_1 + \cdots + L_G$. 
	For the sake of simplicity, we suppose that the first $R$ outcome features are successively categorized into $G$ communities, i.e., $r_{1, 1} = 1$, $\ldots$, $r_{1, L_1} = L_1$, $r_{2, 1} = L_1 + 1$, $\ldots$, $r_{2, L_2} = L_1 + L_2$, and so forth, $r_{G, 1} = R - L_G + 1$, $\ldots$, $r_{G, L_G} = R$ throughout the paper. 
	
	In addition to the community membership, as illustrated in Figure~\ref{Fig:NMR} (C), the (population version) interconnected community structure also demonstrates that: (1) any two features $r_1$ and $r_2$ within community $g$ are coherently and positively correlated, i.e., $\operatorname{cor}\left(y_{r_1}, y_{r_2} \right)$ is a positive constant depending on $g$ only, where $\phi ^ * (r_1) = \phi ^ * (r_2) = g$; and (2) communities $g$ and $g'$ can further be \emph{interconnected}, denoted by $g \sim g'$: $\operatorname{cor} \left(y_{r}, y_{r'} \right) \neq 0$, depending on $g$ and $g'$ only, where $\phi ^ * (r) = g$ and $\phi ^ * (r') = g'$.
	
	In this paper, we are interested in studying the association between $\textbf{y}_i ^ *$ and the covariate vector $\textbf{x}_{i} \in \mathbb{R} ^ {p \times 1}$ for $i = 1, \ldots, n$, while accounting for the interconnected community structure through the following straightforward simultaneous autoregressive regression model:
	\begin{equation}
		\label{Eq:feature}
		\scriptsize	
		%\label{Eq:location}
		%\begin{split}
		y_{ir} ^ *
		\begin{cases}
			\overset{\text{MAUD}}{=}
			\bm{\beta}_{r} ^ {*, \top} \textbf{x}_i 
			+ \underbrace{\frac{1}{L_g - 1} \sum_{r' \neq r: \phi ^ * (r') = \phi ^ *(r) = g}
				\rho_{gg} \left(y_{ir'} ^ * - \bm{\beta}_{r'} ^ {*, \top} \textbf{x}_i \right)}_{\clap{within community $g \geq 1$~}}  \\
			+ \sum_{g': g' \sim g}
			\underbrace{\frac{1}{\sqrt{\left(L_g - 1\right)\left(L_{g'} - 1 \right)}}
				\sum_{r'': \phi ^ * (r'') = g'}
				\rho_{gg'} \left(y_{ir''} ^ * - \bm{\beta}_{r''} ^ {*, \top}  \textbf{x}_i \right)}_{\clap{between communities $g$ and $g' \geq 1$~}}
			+ \epsilon_{ir} ^ *, & \text{ if } r \overset{\text{i.e.}, \phi ^ * (r) \neq 0}{=} 1, \ldots, R \\ 
			\overset{\text{general linear model}}{=}
			\bm{\beta}_{r} ^ {*, \top} \textbf{x}_i + \epsilon_{ir} ^ *, & \text{ if } r \overset{\text{i.e.}, \phi ^ * (r) = 0}{=} R + 1, \ldots, S
		\end{cases},
		%\end{split}
	\end{equation}
	where we remark that model~\eqref{Eq:feature} will be reformulated into our proposed model MAUD (see the next section) for the features in communities: $r = 1, \ldots, R$, or equivalently, $\phi ^ * (r) \neq 0$; while model~\eqref{Eq:feature} simplifies to a general linear model for the singleton features: $r = R + 1, \ldots, S$, or equivalently, $\phi ^ * (r) = 0$.
	Here, $y_{ir} ^ *$ denotes a measurement of $r$th feature in outcomes (e.g., a biomarker),
	$\bm{\beta}_{r} ^ *$ denotes the $p$-dimensional feature-specific regression coefficient vector,
	$\top$ denotes the transpose of a vector or matrix,
	$\textbf{x}_{i}$ denotes the $p$-dimensional covariate vector across all features for the $i$th participant (e.g., intercept, age, and sex),
	$\rho_{gg}$ denotes the dependence parameter within community $g$, 
	$\rho_{gg'}$ denotes the dependence parameter between communities $g$ and $g'$ with $\rho_{gg'} = \rho_{g'g}$, 
	$\epsilon_{ir} ^ *$ denotes an error following $N\left(0, \omega_{r} ^ * \right)$ with a feature-specific variation $\omega_{r} ^ * > 0$ for $i = 1, \ldots, n$ and $r = 1, \ldots, R, R + 1, \ldots, S$.

	Model~\eqref{Eq:feature} extends conventional autoregressive models that are popularly adopted in multivariate analysis and neuroimaging studies \citep{Bowman2005, DeradoBowmanKilts2010, RiskMattesonSpreng2016, LeeChenKochunov2023} by incorporating a generalized framework that accounts for community-wise dependence. 
	In the scenario where all dependence parameters, both within and between communities, are set to $0$, model~\eqref{Eq:feature} converges to a standard general linear model \citep{WorsleyFriston1995, FristonHolmesPoline1995}.
	Furthermore, if the dependence parameters within communities $\rho_{gg} \neq 0$ while those between communities $\rho_{gg'} = 0$ for every $g \neq g'$, model~\eqref{Eq:feature} resembles other existing models found in the literature \citep{Bowman2005, DeradoBowmanKilts2010, LeeChenKochunov2023}.

	Statistical inference about regression parameters (or the dependence parameters) in the above conventional regression models presents a threefold challenge. 
	Firstly, it frequently necessitates iterative algorithms to separately estimate regression and dependence parameters, e.g., the feasible generalized least squares (FGLS) method. 
	Secondly, explicit variance estimation or the finite-sample properties of an FGLS estimator are generally intractable \citep{Hayashi2011}. 
	Meanwhile, the Fisher scoring method can be computationally expensive, even though the dimension $R$ is small \citep{Bowman2005, DeradoBowmanKilts2010}.
	Thirdly, estimating a high-dimensional covariance matrix, especially one subject to certain constraints, is essentially difficult. 
	We will present how to tackle this challenge in the proposed estimation and inference procedures for a more comprehensive regression model, which will be detailed in the following section.
	
	\subsection{Modeling Dependence by The MAUD}
	
	\label{Subsec:maud}
	
	%Given the intricate interconnected community structure, it is necessary to employ a matrix representation of~\eqref{Eq:feature} for parameter estimation. 
	%Without loss of generality, we concentrate on the features in communities only (i.e., the first $R$ features, as illustrated in Figure~\ref{Fig:NMR} (C)), as the features not in any communities can be estimated separately.
	Instead of following the traditional autoregression scheme, in this section, we aim to conduct separate statistical inferences about the parameters in model~\eqref{Eq:feature}.
	Specifically, on the one hand, we derive the MAUD from model~\eqref{Eq:feature} using the matrix representation of the $R$ features in communities: $r = 1, \ldots, R$ in model~\eqref{Eq:feature} or the first $R$ features, as illustrated in Figure~\ref{Fig:NMR} (C). 
	On the other hand, we separately address the general linear regression model derived from model~\eqref{Eq:feature} using the remaining singleton features (those not belonging to any communities: $r = R + 1, \ldots, S$ in model~\eqref{Eq:feature} or the $(S - R)$ features outside the black frame, as illustrated in Figure~\ref{Fig:NMR} (B).

	Dropping the notation $*$ (since the singletons are not included: $y_{ir} = y_{ir} ^ *$, $\bm{\beta}_r = \bm{\beta}_r ^ *$, $\epsilon_{ir} = \epsilon_{ir} ^ *$, $\omega_{r} = \omega_{r} ^ *$ for $r = 1, \ldots, R$) and rewriting the first line of model~\eqref{Eq:feature}, we present the MAUD for features in communities below:
	\begin{equation}
		\label{Eq:individual}
		\textbf{y}_i 
		= \begin{pmatrix}
			\bm{\beta}_{1} ^ \top \\ \vdots \\ \bm{\beta}_{R} ^ \top
		\end{pmatrix}_{R \times p} \textbf{x}_i 
		+ \bm{\Upsilon}_{R \times R} \left[ \textbf{y}_i - \begin{pmatrix}
			\bm{\beta}_{1} ^ \top \\ \vdots \\ \bm{\beta}_{R} ^ \top
		\end{pmatrix}_{R \times p} \textbf{x}_i \right]
		+ \bm{\epsilon}_i, \quad
		\bm{\epsilon}_i \overset{\text{i.i.d.}}{\sim} N \left(\textbf{0}_{R \times 1}, \bm{\Sigma}_{\bm{\epsilon}} \right), 
	\end{equation}
	for $i = 1, \ldots, n$, where $\textbf{y}_i = \left(y_{i1}, \ldots, y_{iR} \right) ^ \top$, $\bm{\epsilon}_i = \left(\epsilon_{i1}, \ldots, \epsilon _{iR} \right) ^ \top$, 
	$\bm{\Sigma}_{\bm{\epsilon}} = \operatorname{cov}(\bm{\epsilon}_i)  = \operatorname{diag} \left(\omega_{1}, \ldots, \omega_{R} \right)$, 
	the dependence parameter matrix $\bm{\Upsilon}$ of the autoregressive regression model has a block form $\left(\bm{\Upsilon}_{gg'}\right)$: 
	\begin{alignat*}{5}
		%\bm{\Upsilon} & = \begin{pmatrix}
		%	\bm{\Upsilon}_{11} & \bm{\Upsilon}_{12} & \dots & \bm{\Upsilon}_{1G} \\
		%	\bm{\Upsilon}_{21} & \bm{\Upsilon}_{22} & \dots & \bm{\Upsilon}_{2G} \\
		%	\vdots & \vdots & \ddots & \vdots \\
		%	\bm{\Upsilon}_{G1} & \bm{\Upsilon}_{G2} & \dots & \bm{\Upsilon}_{GG}
		%\end{pmatrix}, \\
		%\label{Eq:upsilon1}
		\bm{\Upsilon}_{gg} & = \gamma_{gg}\left(\textbf{J}_{L_g} - \textbf{I}_{L_g}\right) \in \mathbb{R} ^ {L_g \times L_g},
		\quad && \gamma_{gg} && = \frac{\rho_{gg}}{L_g - 1}, \quad && g' && = g, \\
		%\label{Eq:upsilon2}
		\bm{\Upsilon}_{gg'} & = \gamma_{gg'} \textbf{J}_{L_g \times L_{g'}} \in \mathbb{R} ^ {L_g \times L_{g'}},
		\quad && \gamma_{gg'} && = \gamma_{g'g} = \frac{\rho_{gg'}} {\sqrt{(L_g - 1)(L_{g'} - 1)}}, \quad && g' && \neq g, 
	\end{alignat*}
	$\gamma_{gg'}$ is the dependence parameter $\rho_{gg'}$ scaled by a constant, $\textbf{I}_n$ and $\textbf{J}_n$ denote the $n$ by $n$ identity matrix and matrix of ones, respectively. 
	We further present $\bm{\Upsilon}$ in terms of $\gamma_{gg'}$:
	\begin{equation}
		\label{Eq:Upsilon_AB}
		\bm{\Upsilon}\left(\textbf{A}_{\bm{\Upsilon}}, \textbf{B}_{\bm{\Upsilon}}, \bm{\ell} \right) = \textbf{A}_{\bm{\Upsilon}} \circ \textbf{I}(\bm{\ell}) + \textbf{B}_{\bm{\Upsilon}} \circ \textbf{J}(\bm{\ell}),
		\text{ with } 
		\begin{cases}
			\textbf{A}_{\bm{\Upsilon}} & = \operatorname{diag}\left(- \gamma_{11}, \ldots, - \gamma_{GG}\right) \\
			\textbf{B}_{\bm{\Upsilon}} & = \left(\gamma_{gg'}\right)
		\end{cases},
	\end{equation}  
	where $\textbf{I}(\bm{\ell}) = \operatorname{Bdiag}\left(\textbf{I}_{L_1}, \ldots, \textbf{I}_{L_G}\right)$ is a block diagonal matrix, $\textbf{J}(\bm{\ell}) = \left(\textbf{J}_{L_{g} \times L_{g'}}\right)$ is a partitioned matrix, and $\circ$ is the block Hadamard product resulting in a block diagonal matrix $\textbf{A}_{\bm{\Upsilon}} \circ \textbf{I}(\bm{\ell}) = \operatorname{Bdiag}\left(-\gamma_{11} \textbf{I}_{L_1}, \ldots, -\gamma_{GG} \textbf{I}_{L_G}\right)$ and a partitioned matrix $\textbf{B}_{\bm{\Upsilon}} \circ \textbf{J}(\bm{r}) = \left(\gamma_{kk'} \textbf{J}_{L_{g} \times L_{g'}}\right)$.
	The above matrix $\bm{\Upsilon}(\textbf{A}_{\bm{\Upsilon}}, \textbf{B}_{\bm{\Upsilon}}, \bm{\ell})$ is also known as the \emph{uniform-block} matrix \citep{YangChenChen2024} (see the definition in the \href{Supplementary Material.pdf}{Supplementary Material}).

	Equivalently, we rewrite the MAUD in~\eqref{Eq:individual} into the following form: 
	\begin{equation}
		\label{Eq:maud_ind}
		\textbf{y}_i \sim N \left(\mathfrak{B}_{R \times p} \textbf{x}_i, \bm{\Sigma} \right), \quad
		\bm{\Sigma} = \left(\textbf{I}_R - \bm{\Upsilon} \right) ^ {-1} \bm{\Sigma}_{\bm{\epsilon}} \left(\textbf{I}_R - \bm{\Upsilon} \right) ^ {-1},
	\end{equation}
	for $i = 1, \ldots, n$, 
	where $\mathfrak{B} ^ \top = \left(\bm{\beta}_{1}, \ldots, \bm{\beta}_{R} \right) \in \mathbb{R} ^ {p \times R}$.
	It's worth mentioning that our primary model MAUD, as described in~\eqref{Eq:maud_ind}, exhibits a widely recognized form that can be deduced from models other than the one presented in~\eqref{Eq:feature}.
	For the sake of easy presentation, we set $\bm{\Sigma}_{\bm{\epsilon}} = \textbf{I}_{R}$ and present the case of $\bm{\Sigma}_{\bm{\epsilon}} \neq \textbf{I}_{R}$ in the \href{Supplementary Material.pdf}{Supplementary Material}.
	
	Assuming that $\bm{\Sigma}$ in~\eqref{Eq:maud_ind} is positive definite, we define $\bm{\Omega} = \bm{\Sigma} ^ {- 1}$ as the precision matrix.
	Leveraging the block Hadamard product, as detailed in the \href{Supplementary Material.pdf}{Supplementary Material}, we can reparametrize $\bm{\Sigma}$ and $\bm{\Omega}$ using $\gamma_{gg'}$ in closed forms, allowing the elements of $\bm{\Omega}$ in the likelihood function to be substituted with $\gamma_{gg'}$. 
	Consequently, $\gamma_{gg'}$ can be directly estimated by maximizing the likelihood function. 
	We resort to Corollary~\ref{Cor:SigmaUB} for the derivation.
	
	%\begin{cor}[parameterize $\gamma_{gg'}$ into $\bm{\Sigma}$]
	\begin{cor}[express $\bm{\Sigma}$ and $\bm{\Omega}$ in terms of $\gamma_{gg'}$ or $\rho_{gg'}$]
		\label{Cor:SigmaUB}
		$\bm{\Omega}$ and $\bm{\Sigma}$ can be expressed as: 
		$\bm{\Omega} = \bm{\Omega} \left(\gamma_{11}, \ldots, \gamma_{GG} \right) = \bm{\Omega}\left(\textbf{A}_{\bm{\Omega}}, \textbf{B}_{\bm{\Omega}}, \bm{\ell} \right)$, $\bm{\Sigma} = \bm{\Sigma} \left(\gamma_{11}, \ldots, \gamma_{GG} \right) = \bm{\Sigma}\left(\textbf{A}_{\bm{\Sigma}}, \textbf{B}_{\bm{\Sigma}}, \bm{\ell} \right)$,
		where
		\begin{align}
			\label{Eq:Omega_AB}
			\bm{\Omega} & = \textbf{A}_{\bm{\Omega}} \circ \textbf{I}(\bm{\ell}) + \textbf{B}_{\bm{\Omega}} \circ \textbf{J}(\bm{\ell}), 
			\text{ with } 
			\begin{cases}
				\textbf{A}_{\bm{\Omega}} = \left(\textbf{I}_{G} - \textbf{A}_{\bm{\Upsilon}} \right) ^ 2 \\
				\textbf{B}_{\bm{\Omega}} = - 2 \textbf{B}_{\bm{\Upsilon}} + \textbf{A}_{\bm{\Upsilon}}   \textbf{B}_{\bm{\Upsilon}} + \textbf{B}_{\bm{\Upsilon}}  \textbf{A}_{\bm{\Upsilon}} + \textbf{B}_{\bm{\Upsilon}}  \textbf{L}  \textbf{B}_{\bm{\Upsilon}}
			\end{cases}, \\
			%%%
			\label{Eq:Sigma_AB}
			\bm{\Sigma} & = \textbf{A}_{\bm{\Sigma}} \circ \textbf{I}(\bm{\ell}) + \textbf{B}_{\bm{\Sigma}} \circ \textbf{J}(\bm{\ell}),
			\text{ with } 
			\begin{cases}
				\textbf{A}_{\bm{\Sigma}} = \textbf{A}_{\bm{\Omega}} ^ {- 1} \\
				\textbf{B}_{\bm{\Sigma}} = - \bm{\Delta}_{\bm{\Omega}} ^ {- 1}  \textbf{B}_{\bm{\Omega}}  \textbf{A}_{\bm{\Omega}} ^ {- 1}
			\end{cases},
		\end{align}
		$\textbf{A}_{\bm{\Upsilon}} = \textbf{A}_{\bm{\Upsilon}} \left(\gamma_{11}, \ldots, \gamma_{GG} \right)$, $\textbf{B}_{\bm{\Upsilon}} = \textbf{B}_{\bm{\Upsilon}} \left(\gamma_{11}, \ldots, \gamma_{GG} \right)$ are defined in~\eqref{Eq:Upsilon_AB}, 
		$\textbf{A}_{\bm{\Sigma}} = \textbf{A}_{\bm{\Sigma}} \left(\gamma_{11}, \ldots, \gamma_{GG} \right)$, $\textbf{A}_{\bm{\Omega}} = \textbf{A}_{\bm{\Omega}} \left(\gamma_{11}, \ldots, \gamma_{GG} \right)$ are diagonal matrices, 
		$\textbf{B}_{\bm{\Sigma}} = \textbf{B}_{\bm{\Sigma}} \left(\gamma_{11}, \ldots, \gamma_{GG} \right)$, $\textbf{B}_{\bm{\Omega}} = \textbf{B}_{\bm{\Omega}} \left(\gamma_{11}, \ldots, \gamma_{GG} \right)$ are symmetric matrices, 
		$\textbf{L} = \operatorname{diag}\left(L_1, \ldots, L_G \right)$, 
		and $\bm{\Delta}_{\bm{\Omega}} = \bm{\Delta}_{\bm{\Omega}} \left(\gamma_{11}, \ldots, \gamma_{GG} \right) = \textbf{A}_{\bm{\Omega}} + \textbf{B}_{\bm{\Omega}} \textbf{L}$.
	\end{cor}
	
	%Corollary~\ref{Cor:SigmaUB} establishes closed-form representations of $\bm{\Omega}$ and $\bm{\Sigma}$ in terms of $\bm{\Upsilon}$ (or $\gamma_{gg'}$).
	%These representations are derived by leveraging several properties of the block Hadamard product, as detailed in the \href{Supplementary Material.pdf}{Supplementary Material}. 

	\subsection{Estimating the MAUD parameters}
	
	\label{Subsec:estimation}
	
	The goal of this section is to estimate the vector of scaled dependence parameters, denoted by $\bm{\gamma} = \left(\gamma_{11}, \ldots, \gamma_{1G}, \gamma_{22}, \ldots, \gamma_{2G}, \ldots, \gamma_{GG}\right) ^ \top \in \mathbb{R} ^ {G (G + 1)/2 \times 1}$, and the regression coefficient vector, denoted by $\bm{\beta} = \operatorname{vec}\left(\mathfrak{B} ^ \top \right) \in \mathbb{R} ^ {(Rp) \times 1}$, simultaneously in the MAUD in~\eqref{Eq:maud_ind}, where $\operatorname{vec}(\cdot)$ denotes the vector of columns of a matrix \citep{HendersonSearle1979}.
	We consider the following multivariate normal distribution for the MAUD:
	\begin{equation*}
		%\label{Eq:maud_vec}
		\textbf{y}_{(nR) \times 1} \sim N \left(\textbf{x}_{(nR) \times (Rp)} \bm{\beta}_{(Rp) \times 1}, \textbf{I}_n \otimes \bm{\Sigma}(\bm{\gamma})\right), 
	\end{equation*}
	where $\textbf{y} = \left(\textbf{y}_1 ^ \top, \ldots, \textbf{y}_n ^ \top \right) ^ \top$, $\textbf{x} = \left(\left(\textbf{I}_R \otimes \textbf{x}_1 ^ \top\right) ^ \top, \dots, \left(\textbf{I}_R \otimes \textbf{x}_n ^ \top \right) ^ \top \right) ^ \top$, and $\otimes$ denotes the Kronecker product.
	
	\textbf{\emph{Parameter estimation by FGLS.}} 
	In statistical literature, FGLS is frequently employed for estimating parameters when correlated multivariate outcomes have an unknown covariance matrix.
	The likelihood function is 
	\begin{equation} 
		\label{Eq:logl1}
		\ell_n \left(\bm{\beta}, \bm{\gamma}; \textbf{x}, \textbf{y} \right)
		= \frac{n}{2}
		\left [
		- R \log(2 \pi)
		+ \log \det \left(\bm{\Omega}\right)
		- \operatorname{tr} \left(\textbf{S} \bm{\Omega}\right)
		\right ],
	\end{equation}
	where $\bm{\Omega} = \bm{\Omega}(\bm{\gamma})$ defined in~\eqref{Eq:Omega_AB}, 
	$\textbf{S} = \textbf{S}(\bm{\beta}) = n ^ {-1} \sum_{i=1}^{n}\left(\textbf{y}_i - \mathfrak{B}\textbf{x}_i \right) \left(\textbf{y}_i - \mathfrak{B}\textbf{x}_i\right) ^ \top \in \mathbb{R} ^ {R \times R}$ is the residual matrix.
	
	Typically, we adopt the FGLS approach to estimate $\bm{\beta}$ and $\bm{\gamma}$ iteratively: at iteration $t \geq 1$, the FGLS estimators of $\bm{\beta}$ and $\bm{\gamma}$ are straightforward by
	\begin{equation*}
		%\label{Eq:FGLS}
		\widehat{\bm{\beta}} ^ {(t)}
		%= \left(\textbf{x} ^ \top \textbf{x} \right) ^ {-1} \textbf{x} ^ \top \textbf{y} \in \mathbb{R} ^ {(Rp) \times 1}.
		= \left\{\textbf{x} ^ \top \left[\textbf{I}_n \otimes \widehat{\bm{\Omega}} ^ {(t - 1)} \right] \textbf{x}\right\} ^ {-1} \textbf{x} ^ \top \left(\textbf{I}_n \otimes \widehat{\bm{\Omega}} ^ {(t - 1)} \right) \textbf{y}, \quad %\in \mathbb{R} ^ {(Rp) \times 1},
		\widehat{\bm{\gamma}} ^ {(t)}
		\in \argmax \limits_{\bm{\gamma} \in \Theta_{\bm{\gamma}}}
		\ell_n\left(\widehat{\bm{\beta}} ^ {(t)}, \bm{\gamma}; \textbf{x}, \textbf{y}\right),
	\end{equation*}
	where $\Theta_{\bm{\gamma}} \subset \mathbb{R} ^ {G(G+1)/2 \times 1}$ is the parameter space of $\bm{\gamma}$, $\widehat{\bm{\Omega}} ^ {(t-1)} = \widehat{\bm{\Omega}}\left(\widehat{\bm{\gamma}} ^ {(t - 1)} \right)$ and $\widehat{\bm{\gamma}} ^ {(t-1)}$ are the estimators of $\bm{\Omega}$ and $\bm{\gamma}$ at the $(t-1)$th iteration, respectively.
	Initially, $\widehat{\bm{\beta}} ^ {(0)}$ is the OLS estimator. 
	Plugging $\widehat{\bm{\beta}} ^ {(t)}$, we further compute $\widehat{\bm{\gamma}} ^ {(t)}$ by updating $\textbf{S} ^ {(t)} = n ^ {-1} \sum_{i=1}^{n}\left(\textbf{y}_i - \widehat{\mathfrak{B}} ^ {(t)} \textbf{x}_i \right) \left(\textbf{y}_i - \widehat{\mathfrak{B}} ^ {(t)} \textbf{x}_i\right) ^ \top$ and maximizing~\eqref{Eq:logl1}.
	
	As mentioned earlier, the estimation procedure for $\bm{\gamma}$ through iterative optimization of ~\eqref{Eq:logl1} poses challenges due to the inherent difficulty in estimating large covariance or precision matrices \citep{FanLiaoLiu2016, CaiRenZhou2016}.
	Conversely, employing advanced methods for large covariance or precision matrix estimation (e.g., shrinkage and thresholding approaches) is not suitable as they may result in biased estimates of dependence parameters in the model.
	This is due to the constraints imposed, where the diagonals of each $\bm{\Upsilon}_{gg}$ are enforced to be $0$'s.
	
	\textbf{\emph{Parameter estimation by MAUD.}}
	In our estimation procedure, Corollary~\ref{Cor:SigmaUB} plays a critical role. 
	Specifically, plugging $\widehat{\bm{\beta}} ^ {(t)}$, we equivalently rewrite the likelihood function in~\eqref{Eq:logl1}:
	\begin{align}
		& \ell_n \left(\widehat{\bm{\beta}} ^ {(t)}, \bm{\gamma}; \textbf{x}, \textbf{y} \right)
		=
		\frac{n}{2} \bigg\{
		- R \log(2 \pi)
		+ \sum_{g = 1}^{G} \left(L_g - 1\right) \log \left(a_{\bm{\Omega}, gg} \right)
		+ \log \det \left(\bm{\Delta}_{\bm{\Omega}}\right) \nonumber \\
		& - \operatorname{sum} \left[\textbf{A}_{\bm{\Omega}} \odot \operatorname{diag}\left(\operatorname{tr}\left(\textbf{S}_{11} ^ {(t)} \right), \ldots, \operatorname{tr}\left(\textbf{S}_{GG} ^ {(t)} \right)\right) 
		+ \textbf{B}_{\bm{\Omega}} \odot \left(\operatorname{sum}\left(\textbf{S}_{gg'} ^ {(t)} \right)\right) \right] \bigg\}, \label{Eq:logl2}
	\end{align}
	where $a_{\bm{\Omega}, gg}$ is the $g$th diagonal element of $\textbf{A}_{\bm{\Omega}} = \textbf{A}_{\bm{\Omega}}\left(\bm{\gamma}\right) \in \mathbb{R} ^ {G \times G}$ defined in~\eqref{Eq:Omega_AB}; 
	$\textbf{B}_{\bm{\Omega}} = \textbf{B}_{\bm{\Omega}}\left(\bm{\gamma}\right) \in \mathbb{R} ^ {G \times G}$ defined in~\eqref{Eq:Omega_AB};
	$\bm{\Delta}_{\bm{\Omega}} = \bm{\Delta}_{\bm{\Omega}}\left(\bm{\gamma}\right) \in \mathbb{R} ^ {G \times G}$ defined in Corollary~\ref{Cor:SigmaUB}; 
	$\left(\textbf{S}_{gg'} ^ {(t)} \right) = \left(\textbf{S}_{gg'} ^ {(t)} \right)_{g, g' = 1} ^ {G} $ is the $G$ by $G$ partitioned matrix of $\textbf{S} ^ {(t)}$ satisfying $\textbf{S}_{gg'} ^ {(t)} \in \mathbb{R} ^ {L_g \times L_{g'}}$,
	$\operatorname{sum}(\cdot)$ denotes the sum of all elements of a matrix; and $\odot$ denotes the (entry-wise) Hadamard product.
	Contrasted with the log-likelihood function in~\eqref{Eq:logl1}, the alternative form presented in~\eqref{Eq:logl2} significantly reduces the number of parameters, i.e., from $R(R + 1)/2$ for $\bm{\Omega}$ to $G(G + 1)/2$ for $\bm{\gamma}$, e.g., from $14,535$ to $15$ for the NMR data with $R = 170$ and $G = 5$.
	
	In addition to simplifying matrix calculations in the log-likelihood function, the following theorems establish that we can obtain the estimates without relying on the iterative relationship in the FGLS estimators.
	Before proceeding to the theorems, we introduce the following estimators:
	\begin{equation}
		\label{Eq:OLS}
		\widehat{\bm{\beta}} = \left(\textbf{x} ^ \top \textbf{x} \right) ^ {-1} \textbf{x} ^ \top \textbf{y}, \quad
		%\label{Eq:MLE}
		\widehat{\bm{\gamma}} \in \argmax \limits_{\bm{\gamma} \in \Theta_{\bm{\gamma}}}
		\ell_n \left(\widehat{\bm{\beta}}, \bm{\gamma}; \textbf{x}, \textbf{y} \right),
	\end{equation}
	where classical algorithms (e.g., the Fisher scoring) can be readily employed to obtain $\widehat{\bm{\gamma}}$ by optimizing the log-likelihood function in~\eqref{Eq:logl2} given $\widehat{\bm{\beta}}$.
	Since the number of covariance parameters $G(G+1)/2$ is often small, the computational load is less expensive (e.g., a few seconds).
	Accordingly, we derive the estimator $\widehat{\bm{\rho}} = \left(\widehat{\rho}_{11}, \ldots, \widehat{\rho}_{GG}\right) ^ \top \in \mathbb{R} ^ {G(G+1)/2 \times 1}$ of $\rho_{gg'}$ in~\eqref{Eq:feature} with $\widehat{\rho}_{gg} = (L_g - 1) \widehat{\gamma}_{gg}$ for $g' = g$ and $\widehat{\rho}_{gg'} = \widehat{\rho}_{g'g} = \sqrt{\left(L_{g} - 1 \right) \left(L_{g'} - 1 \right)} \widehat{\gamma}_{gg'}$ for $g' \neq g$, which completes the MAUD parameter estimation.
	
	\begin{thm}[Finite- and large-sample properties of $\widehat{\bm{\beta}}$]
		\label{Thm:beta}
		Suppose Conditions 1-6 (presented in the \href{Supplementary Material.pdf}{Supplementary Material}) are satisfied. %Conditions~\ref{Con:pd},~\ref{Con:linear_independent},~\ref{Con:sample_size},~\ref{Con:beta_gamma},~\ref{Con:beta_converge}, and~\ref{Con:diagonal_converge} hold.
		Under finite sample size (as $n$ is fixed),
		the OLS estimator $\widehat{\bm{\beta}}$ in~\eqref{Eq:OLS} is identical to both the GLS estimator and the FGLS estimator:
		\begin{equation*}
			\begin{split}
				\widehat{\bm{\beta}}
				%& = \widehat{\bm{\beta}} ^ {(0)} \nonumber \\
				%& = \left(\textbf{x} ^ \top \textbf{x} \right) ^ {-1} \textbf{x} ^ \top \textbf{y} \nonumber \\
				= \left\{\textbf{x} ^ \top \left(\textbf{I}_n \otimes \bm{\Omega} \right) \textbf{x}\right\} ^ {-1} \textbf{x} ^ \top \left(\textbf{I}_n \otimes \bm{\Omega} \right) \textbf{y} \nonumber 
				= \left\{\textbf{x} ^ \top \left(\textbf{I}_n \otimes \widehat{\bm{\Omega}} ^ {(t-1)}\right) \textbf{x}\right\} ^ {-1} \textbf{x} ^ \top \left(\textbf{I}_n \otimes \widehat{\bm{\Omega}} ^ {(t-1)} \right) \textbf{y} \nonumber 
				= \widehat{\bm{\beta}} ^ {(t)},
			\end{split}
		\end{equation*}
		for all $t \geq 1$;
		and $\widehat{\bm{\beta}} \sim N \left(\bm{\beta}, \bm{\Sigma} \left(\textbf{A}_{\bm{\Sigma}}, \textbf{B}_{\bm{\Sigma}}, \bm{\ell} \right) \otimes \left(\sum_{i = 1} ^ {n} \textbf{x}_i \textbf{x}_i ^ \top \right) ^ {-1} \right)$.
		Under large sample size (as $n \to \infty$), 
		$\widehat{\bm{\beta}}$ is (weakly) consistent, asymptotically efficient, asymptotically normally distributed, 
	\end{thm}

	\begin{thm}[Large-sample properties of $\widehat{\bm{\gamma}}$ and $\widehat{\bm{\rho}}$]
		\label{Thm:gamma}
		Suppose Conditions 1-6 are satisfied.
		%Suppose Conditions~\ref{Con:pd},~\ref{Con:linear_independent},~\ref{Con:sample_size},~\ref{Con:beta_gamma},~\ref{Con:beta_converge}, and~\ref{Con:diagonal_converge} hold.
		Then,  
		$\widehat{\bm{\gamma}}$ in~\eqref{Eq:OLS} and $\widehat{\bm{\rho}}$ are the unique ML estimators.
		As $n \to \infty$, they are (weakly) consistent, asymptotically efficient, asymptotically normally distributed.
	\end{thm}

	We denote $\bm{\theta} = \left(\bm{\beta} ^ \top, \bm{\gamma} ^ \top \right) ^ \top \in \mathbb{R} ^ {[(R p) + G(G + 1)/2] \times 1}$.
	The following theorem establishes the equivalence between the joint estimation of $\bm{\theta}$ and the separate estimation of $\bm{\beta}$ and $\bm{\gamma}$, the independence between $\widehat{\bm{\beta}}$ and $\widehat{\bm{\gamma}}$, as well as the Fisher information matrix. 
	
	\begin{thm}[Large-sample properties of $\widehat{\bm{\theta}}$]
		\label{Thm:theta}
		Suppose Conditions 1-6 are satisfied.
		%Suppose Conditions~\ref{Con:pd},~\ref{Con:linear_independent},~\ref{Con:sample_size},~\ref{Con:beta_gamma},~\ref{Con:beta_converge}, and~\ref{Con:diagonal_converge} hold.
		Then, 
		$\widehat{\bm{\theta}} = \left(\widehat{\bm{\beta}} ^ \top, \widehat{\bm{\gamma}} ^ \top\right) ^ \top$ satisfies the score equation with respect to $\bm{\theta}$, i.e., 
		\begin{equation*}
			\frac{\partial }{\partial \bm{\theta}} \ell_n \left(\bm{\theta}; \textbf{x}, \textbf{y} \right) \bigg |_{\bm{\theta} = \widehat{\bm{\theta}}} = \textbf{0}_{[(R p) + G(G + 1)/2] \times 1}.
		\end{equation*}
		In addition, the Fisher information matrix of the log-likelihood function $\ell_n\left(\bm{\theta}; \textbf{x}, \textbf{y} \right)$ is $\bm{\Psi} = \operatorname{Bdiag}\left(\bm{\Psi}_{\bm{\beta}}, \bm{\Psi}_{\bm{\gamma}} \right)$ 
		%
		%\begin{equation*}
		%	\bm{\Psi} = \begin{pmatrix}
		%		\bm{\Psi}_{\bm{\beta}} & \textbf{0}_{(Rp) \times [G(G+1)/2]} \\ \textbf{0}_{[G(G + 1) / 2] \times (R p)} & \bm{\Psi}_{\bm{\gamma}}
		%	\end{pmatrix},
		%\end{equation*}
		%and it 
		is positive definite, 
		with 
		\begin{equation*}
			\begin{split}
				\bm{\Psi}_{\bm{\beta}} 
				& = \textbf{x} ^ \top \left[\textbf{I}_n \otimes \bm{\Omega}\left(\textbf{A}_{\bm{\Omega}}, \textbf{B}_{\bm{\Omega}}, \bm{\ell} \right)\right] \textbf{x}
				= \bm{\Omega}\left(\textbf{A}_{\bm{\Omega}}, \textbf{B}_{\bm{\Omega}}, \bm{\ell} \right) \otimes \left(\sum_{i = 1} ^ {n} \textbf{x}_i \textbf{x}_i ^ \top \right) \in \mathbb{R} ^ {(Rp) \times (Rp)}, \\
				%%%%
				\bm{\Psi}_{\bm{\gamma}} & = \left(\psi_{jj'} ^ {(\bm{\gamma})}\right), \quad 
				\psi_{jj'} ^ {(\bm{\gamma})}
				= \frac{n}{2} \operatorname{tr} \left\{
				\left[\frac{\partial \bm{\Omega} \left(\textbf{A}_{\bm{\Omega}}, \textbf{B}_{\bm{\Omega}}, \bm{\ell} \right)}{\partial \bm{\gamma}_j}\right] 
				\bm{\Sigma} \left(\textbf{A}_{\bm{\Sigma}}, \textbf{B}_{\bm{\Sigma}}, \bm{\ell} \right)
				\left[\frac{\partial \bm{\Omega} \left(\textbf{A}_{\bm{\Omega}}, \textbf{B}_{\bm{\Omega}}, \bm{\ell} \right)}{\partial \bm{\gamma}_{j'}}\right] 
				\bm{\Sigma} \left(\textbf{A}_{\bm{\Sigma}}, \textbf{B}_{\bm{\Sigma}}, \bm{\ell} \right)
				\right\},
			\end{split}
		\end{equation*}
		$\bm{\gamma}_j \in \{\gamma_{11}, \ldots, \gamma_{1G}, \ldots, \gamma_{2G}, \ldots, \gamma_{GG}\}$ denotes the $j$th component of $\bm{\gamma}$ for $j = 1, \ldots, G(G + 1) / 2$,
		and $\psi_{jj'} ^ {(\bm{\gamma})}$ is simplified to a closed-form expression in the \href{Supplementary Material.pdf}{Supplementary Material}.
	\end{thm}
	
	\subsection{Conducting inference about the MAUD parameters}
	
	\label{Subsec:inference}
	
	In this section, we conduct statistical inference methods for both $\bm{\beta}$ and $\bm{\gamma}$ based on the closed-form exact covariance matrix of $\widehat{\bm{\beta}}$ and the closed-form asymptotic covariance matrix of $\widehat{\bm{\gamma}}$. %to access the statistical significance of the regression coefficients and to determine whether the dependence within and between groups can be considered null, respectively.
	
	\textbf{\emph{Inference about $\bm{\beta}$.}}
	We conduct statistical tests for $\bm{\beta}$ using the estimator presented in~\eqref{Eq:OLS}.
	Without loss of generality, we are simultaneously testing $Rp$ covariate-wise hypotheses:
	\begin{equation}
		\label{Eq:H01q}
		H_{0, r, q}: \beta_{r} ^ {(q)} = 0 \quad \text{ versus } \quad 
		H_{1, r, q}: \beta_{r} ^ {(q)} \neq 0, \quad q = 1, \ldots, p, r = 1, \ldots, R,
	\end{equation}
	where $\beta_{r} ^ {(q)}$ is the $q$th component of the $r$th regression coefficient vector, i.e., $\bm{\beta}_{r} = \left(\beta_{r} ^ {(1)}, \ldots, \beta_{r} ^ {(p)} \right) ^ \top$.
	%In general, in~\eqref{Eq:H01q}, $0$ can be replaced with a pre-determined vector and the two-sided hypothesis can be replaced with a one-sided hypothesis.
	
	An exact covariance matrix of $\widehat{\bm{\beta}}$ is available from Theorem~\ref{Thm:beta}:
	\begin{equation}
		\label{Eq:cov_beta}
		\bm{\Sigma}_{\widehat{\bm{\beta}}}
		= \operatorname{cov}\left(\widehat{\bm{\beta}}\right)
		= \bm{\Sigma} \left(\textbf{A}_{\bm{\Sigma}}, \textbf{B}_{\bm{\Sigma}}, \bm{\ell} \right) \otimes \left(\sum_{i = 1} ^ {n} \textbf{x}_i \textbf{x}_i ^ \top \right) ^ {-1} \in \mathbb{R} ^ {(R p) \times (R p)}.
	\end{equation}
	%\citet{LeeChenKochunov2023} proposed a new test statistics for the covariate-wise hypotheses with variance-reduced technique. 
	In general, given a pre-determined matrix $\textbf{C} ^ * \in \mathbb{R} ^ {s \times (Rp)}$ with a full rank, we can test a secondary parameter (SP) $\bm{\varrho} ^ * = \textbf{C} ^ * \bm{\beta} \in \mathbb{R} ^ {s \times 1}$:
	\begin{equation*}
		%\label{Eq:secondary}
		H_{0, \text{SP}}: \bm{\varrho} ^ * = \bm{\varrho}_0 ^ * \quad
		\text{versus} \quad
		H_{1, \text{SP}}: \bm{\varrho} ^ * \neq \bm{\varrho}_0 ^ *.
	\end{equation*} 
	We construct a Wald-type test statistic using the estimator $\widehat{\bm{\varrho}} ^ * = \textbf{C} ^ * \widehat{\bm{\beta}}$ that follows a multivariate normal distribution with mean $\textbf{C} ^ * \bm{\beta} \in \mathbb{R} ^ {s \times 1}$ and covariance matrix $\textbf{C} ^ * \bm{\Sigma}_{\widehat{\bm{\beta}}} \textbf{C} ^ {*, \top} \in \mathbb{R} ^ {s \times s}$.
	
	\textbf{\emph{Inference about $\bm{\gamma}$.}} 
	By applying Theorem~\ref{Thm:theta} and utilizing the log-likelihood function in~\eqref{Eq:logl2}, we can compute the observed information matrix to estimate the asymptotic covariance matrix of $\widehat{\bm{\gamma}}$, i.e., $\operatorname{cov}\left(\widehat{\bm{\gamma}}\right) \overset{\text{asy.}}{=} \widehat{\bm{\Psi}}_{\bm{\gamma}} ^ {-1}$ or $\sqrt{n}\left(\widehat{\bm{\gamma}} - \bm{\gamma}\right) \to N \left(\textbf{0}_{G(G + 1) / 2 \times 1}, n \widehat{\bm{\Psi}}_{\bm{\gamma}} ^ {-1} \right)$ in distribution, where we replace the truths with their estimates in $\bm{\Psi}_{\bm{\gamma}}$ to obtain $\widehat{\bm{\Psi}}_{\bm{\gamma}}$.
	Subsequently, we perform $G (G + 1) / 2$ Wald-type tests to evaluate the null and alternative hypotheses:
	\begin{equation}
		\label{Eq:H01gg'}
		H_{0, gg'}: \gamma_{gg'} = 0 \quad \text{ versus } \quad H_{1, gg'}: \gamma_{gg'} \neq 0, 
		\quad g \leq g' = 1, \ldots, G.
	\end{equation}
	
	Multiple testing correction procedures can be further performed to account for multiplicity (e.g., false discovery rate, FDR) and dependence (e.g., the whitening transform) of the simultaneous tests \citep{BenjaminiHochberg1995, LeekStorey2008, Jin2012}.

	%If the test statistics are independent, or their joint distribution exhibits property of positive regression dependence on subsets (PRDS) \citep{BenjaminiYekutieli2001}, then a standard BH procedure is appropriate for multiple testing.
	%However, if the asymptotic covariance matrix of the joint distribution of $\widehat{\bm{\gamma}}$ is not an identity matrix, we can use a modified version of the BH procedure known as the Benjamini--Yekutieli (BY) procedure to control the FDR. 
	%It is important to note that the BY procedure is conservative and may result in a loss of power \citep{BenjaminiYekutieli2001}.
	%We will discuss the correlation case later.
	
	\section{Simulation Studies}
	
	\label{Sec:simulation}
		
	In this section, we conduct Monte Carlo simulation studies to evaluate the performance of MAUD in terms of parameter estimation and statistical inference about both dependence parameters and regression coefficients.
	We also compare its performance with those of competing linear regression models including the Autoregressive model for Multivariate Block-Diagonal outcomes \citep[AMBD;][]{LeeChenKochunov2023}, the Mixed Model for Repeated Measures \citep[MMRM;][]{RBoveDedicKelkhoff2022}, and the GeneraL Multivariate linear regression Model \citep[GLMM;][]{WorsleyFriston1995}.
	We perform sensitivity analysis for the MAUD under model misspecification when the underlying covariance matrix does not have the interconnected community structure.
	   
	\subsection{Synthetic data}
	
	\label{Subsec:data}
	
	We generate the multivariate outcome features $\textbf{y}_i$ from $N \left(\mathfrak{B} \textbf{x}_i, \bm{\Sigma} \right)$, where $i = 1, \ldots, n$.
	We firstly specify the mean vectors by sampling $\textbf{x}_i$ from a standard normal distribution and defining the regression coefficients $\mathfrak{B} ^ \top = \left(\bm{\beta}_1, \ldots, \bm{\beta}_R \right)_{p \times R}$ with $p = 2$.
	Specifically, we set $\textbf{x}_i = (x_{i1}, x_{i2}) ^ \top$ with $x_{i1} = 1$ and $x_{i2} \sim N(0, 1)$.
	We set $\bm{\beta}_r = \left(\beta_{r} ^ {(1)}, \beta_{r} ^ {(2)} \right) ^ \top$ with $\beta_{r} ^ {(1)} \neq \beta_r ^ {(2)}$ and within each community, we set $30\%$ regression coefficients $\bm{\beta}_r$ to be nonzero, while the remaining regression coefficients are set to be $\textbf{0}_{2 \times 1}$.
	The nonzero regression coefficients $\bm{\beta}_r$ can be different across communities. 
	We explore various configurations with $n$, $G$, $R$, $\bm{\ell}$, and covariance matrices $\bm{\Sigma} = \bm{\Sigma} \left(\textbf{A}_{\bm{\Sigma}}, \textbf{B}_{\bm{\Sigma}}, \bm{\ell} \right)$.
	Specifically, we set $n \in \{100, 200, 300\}$ and fix $G = 3$, $R = 100$, $\bm{\ell} = (30, 40, 60) ^ \top$, and
	\begin{equation*}
		\textbf{A}_{\bm{\Upsilon}}
		= \begin{pmatrix}
			-0.40 & & \\ & -0.19 & \\ & & 0.64
		\end{pmatrix}, \quad
		\textbf{B}_{\bm{\Upsilon}}
		= \begin{pmatrix}
			0.40 & 0.01 & -0.51 \\
			& 0.19 & -0.91 \\
			& & -0.64
		\end{pmatrix}
	\end{equation*}
	to evaluate the finite-sample performance of $\widehat{\bm{\gamma}}$.
	Additionally, we fix $n = 50$ and set $G \in \{3, 4\}$, $R \in \{100, 150, 200\}$, various $\bm{\ell}$ (see Figure~\ref{Fig:study_2}), various $\textbf{A}_{\bm{\Upsilon}}$ and $\textbf{B}_{\bm{\Upsilon}}$, to evaluate the performance of hypothesis testing using $\widehat{\bm{\beta}}$.
	We conduct $1000$ replicates for each setting.
	
	\subsection{Evaluation of scaled dependence parameter estimation by MAUD}
	
	\label{Subsec:study_1}
	
	For each replicated dataset, we employ the MAUD to estimate $\bm{\beta}$ and $\bm{\gamma}$ using~\eqref{Eq:OLS}. 
	We aim to assess the accuracy of the MAUD in estimating the vector of scaled dependence parameters ${\bm{\gamma}}$.  
	To evaluate the performance of estimator $\widehat{\bm{\gamma}}$, we utilize several evaluation metrics, including the average relative bias (denoted as ``bias”), the Monte Carlo standard deviation (denoted as ``MCSD”), the average asymptotic standard error (denoted as ``ASE”), and the $95\%$ coverage probability based on the Wald-type confidence interval (denoted as ``$95\%$ CP”) for each element of $\bm{\gamma}$.
	The estimates of $\bm{\gamma}$ are summarized in Table~\ref{Tab:study_1}.
	
	%%%%%%%%%%%%%%%%%
	%%%% Table 1 %%%%
	%%%%%%%%%%%%%%%%%
	\begin{table}[!h]
		\centering
		\begin{tabular}{cccccc}
			\hline
			$n$ & parameter & bias & MCSD & ASE & $95\%$ CP \\ \hline
			& $\gamma_{11}$ & \phantom{-}1.23 & 1.68 & 1.66 & 88.7 \\
			& $\gamma_{12}$ & \phantom{-}0.49 & 2.56 & 2.51 & 93.6 \\
			100 & $\gamma_{13}$ & -0.71 & 4.29 & 4.15 & 94.7 \\
			& $\gamma_{22}$ & \phantom{-}1.24 & 1.55 & 1.55 & 88.1 \\
			& $\gamma_{23}$ & -1.27 & 5.07 & 4.92 & 94.3 \\
			& $\gamma_{33}$ & \phantom{-}0.07 & 4.67 & 0.41 & 84.8 \\ \hline
			& $\gamma_{11}$ & \phantom{-}0.58 & 1.15 & 1.17 & 92.4 \\
			& $\gamma_{12}$ & \phantom{-}0.20 & 1.74 & 1.75 & 94.9 \\
			200 & $\gamma_{13}$ & -0.36 & 3.02 & 2.90 & 93.5 \\
			& $\gamma_{22}$ & \phantom{-}0.63 & 1.09 & 1.09 & 91.8 \\
			& $\gamma_{23}$ & -0.58 & 3.55 & 3.44 & 94.5 \\
			& $\gamma_{33}$ & \phantom{-}0.18 & 0.29 & 0.29 & 91.0 \\ \hline
			& $\gamma_{11}$ & \phantom{-}0.38 & 0.96 & 0.95 & 92.3 \\
			& $\gamma_{12}$ & \phantom{-}0.14 & 1.40 & 1.42 & 95.3 \\
			300 & $\gamma_{13}$ & -0.24 & 2.37 & 2.36 & 95.1 \\
			& $\gamma_{22}$ & \phantom{-}0.41 & 0.91 & 0.89 & 92.3 \\
			& $\gamma_{23}$ & -0.34 & 2.81 & 2.80 & 95.1 \\
			& $\gamma_{33}$ & \phantom{-}0.12 & 0.24 & 0.24 & 91.6 \\ \hline
		\end{tabular}
		\caption{We present the estimation results ($\times 100$) of $\bm{\gamma}$ for $n = 100, 200, 300$, 
			where 
			``bias" represents the average of estimation bias, 
			``MCSD" signifies the Monte Carlo standard deviation, 
			``ASE" denotes the average asymptotic standard error, 
			``$95\%$ CP" indicates the coverage probability based on a $95\%$ Wald-type confidence interval.}
		\label{Tab:study_1}
	\end{table}
	%%%%%%%%%%%%%%%%%%%%%
	%%% End of Table 1 %%
	%%%%%%%%%%%%%%%%%%%%%
	 
	The results in Table~\ref{Tab:study_1} indicate relatively small biases, especially in comparison to the Monte Carlo standard deviations.
	Additionally, the asymptotic standard errors closely align with the Monte Carlo standard deviations.
	As the sample size grows, the asymptotic standard errors decrease, approaching the Monte Carlo standard deviations.
	Consequently, the coverage probabilities based on $95\%$ Wald-type confidence intervals become closer to the nominal level.
	
	\subsection{Evaluation of regression coefficient inference by MAUD}
	
	\label{Subsec:study_2}
	
\begin{figure}[!h]
		\centering
		\includegraphics[width = 1.0\linewidth]{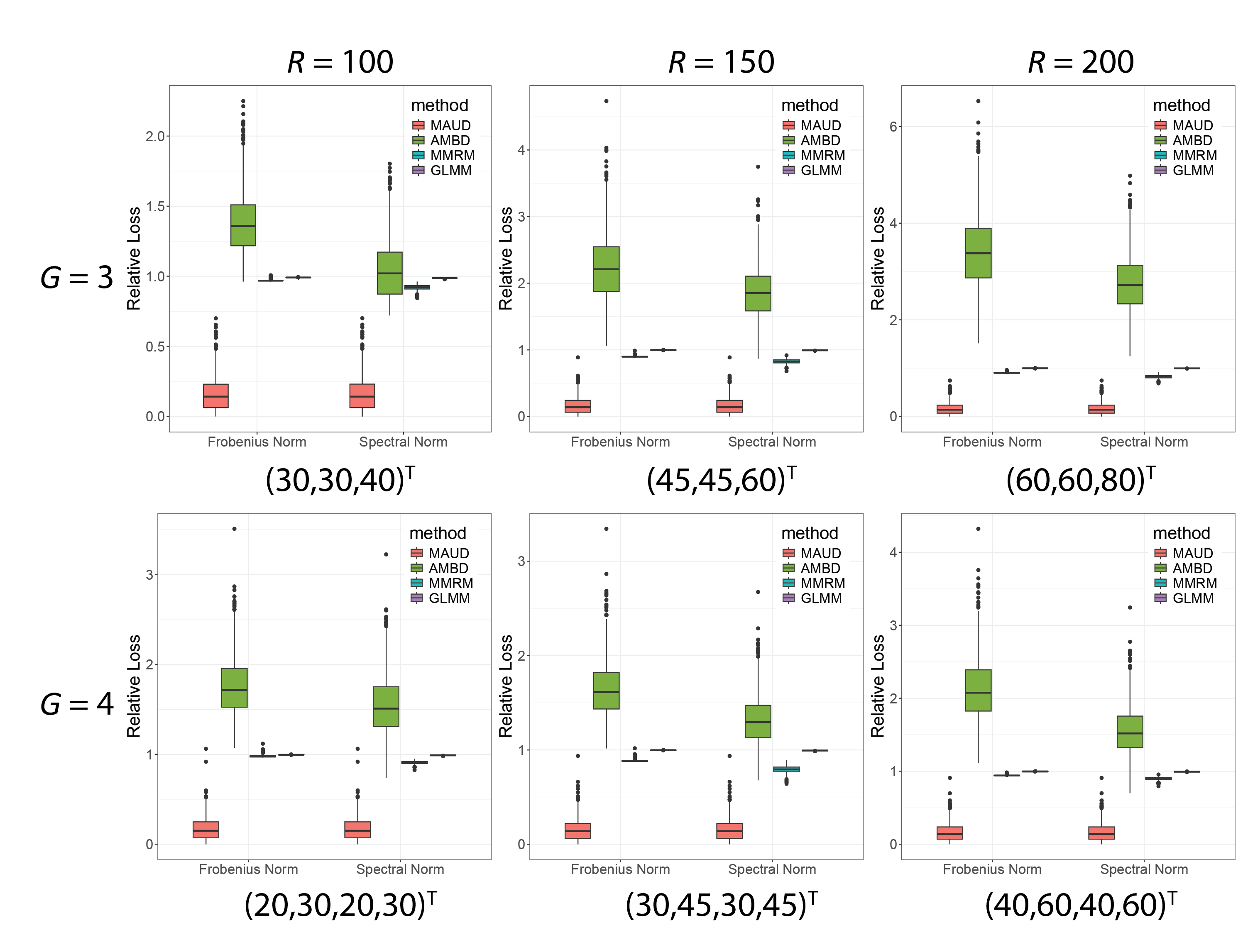} 
		\caption{
			We compare the relative losses of $\bm{\Sigma}_{\widehat{\bm{\beta}}}$ obtained from the MAUD and existing conventional linear regression models.
			The subfigures in the first row demonstrate the relative losses under the setting $n = 50$, $G = 3$, $\bm{\ell} = (30, 30, 40) ^ \top (R = 100)$, $(45, 45, 60) ^ \top (R = 150)$, and $(60, 60, 80) ^ \top (R = 200)$, respectively.
			The subfigures in the second row demonstrate the relative losses under the setting $n = 50$, $G = 4$, $\bm{\ell} = (20, 30, 20, 30) ^ \top (R = 100)$, $(30, 45, 30, 45) ^ \top (R = 150)$, and $(40, 60, 40, 60) ^ \top (R = 200)$, respectively.
		}
		\label{Fig:study_2}
	\end{figure}

	We further conduct an assessment of the statistical inference regarding the regression coefficients using the MAUD and compare it with models of AMBD, MMRM, and GLMM. 
	Our primary focus is on identifying a subset of features associated with the covariates. 
	It is worth noting that a more accurate estimate of $\bm{\Sigma}_{\widehat{\bm{\beta}}}$ in~\eqref{Eq:cov_beta} always implies more precise results in hypothesis testing.  
	To achieve this, we initially analyze the relative loss of $\bm{\Sigma}_{\widehat{\bm{\beta}}}$, defined as $ \Vert \widehat{\bm{\Sigma}}_{\widehat{\bm{\beta}}} - \bm{\Sigma}_{\widehat{\bm{\beta}}} \Vert /  \Vert \bm{\Sigma}_{\widehat{\bm{\beta}}} \Vert$, and generate boxplots of relative losses for all models under various settings in Figure~\ref{Fig:study_2}, utilizing both the Frobenius norm and the spectral norm. 
	Subsequently, we empirically calculate rejection proportions for components of $\bm{\beta}$ in the \href{Supplementary Material.pdf}{Supplementary Material}.

	%%%%%%%%%%%%%%%%%%
	%%%% Figure 3 %%%%
	%%%%%%%%%%%%%%%%%%
	%%%%%%%%%%%%%%%%%%%%%
	%%%End of Figure 3 %%
	%%%%%%%%%%%%%%%%%%%%%
	
	Overall, as illustrated in Figure~\ref{Fig:study_2}, the MAUD exhibits the lowest relative loss among all models across various settings, aligning with the most precise inference results detailed in the \href{Supplementary Material.pdf}{Supplementary Material}. 
	The AMBD demonstrates the highest relative loss due to the inability of its parameter estimation procedure to accommodate cases where the covariance matrix $\bm{\Sigma}$ deviates from the block diagonal structure. 
	Notably, both the MMRM and the GLMM show slightly lower but almost uniformly relative losses compared to the AMBD.
	This can be attributed to our specification of $\bm{\Sigma}$ with an autocorrelation AR(1) structure in the MMRM and a diagonal structure in the GLMM.

	\subsection{Misspecification analysis of MAUD}
	
	\label{Subsec:study_3}
	
	In the last two studies, we operated under the assumption of a known interconnected community structure for the MAUD. 
	However, in practical scenarios, the latent covariance structure may deviate from the interconnected community structure. 
	Considering this, we assess the robustness of the MAUD under potential model misspecifications. 
	
	To evaluate this, we introduce a perturbation term $\textbf{E}_{\sigma} \sim \text{Wishart}\left(R, \sigma \textbf{I}_{R} \right)$, where noise level $\sigma \in \{0, 3, 6, 9\} \times 10 ^ {-2}$ and $\textbf{E}_{\sigma} = \textbf{0}_{R \times R}$ for $\sigma = 0$.
	We generate simulated datasets $1000$ times, incorporating the covariance matrix $\bm{\Sigma} = \bm{\Sigma} \left(\textbf{A}_{\bm{\Sigma}}, \textbf{B}_{\bm{\Sigma}}, \bm{\ell} \right) + \textbf{E}_{\sigma}$ under the setting of $n = 50$, $G = 3$, $R = 100$, and $\bm{\ell} = (30, 30, 40) ^ \top$.
	
\begin{figure}[!h]
		\centering
		\includegraphics[width = 0.5 \linewidth]{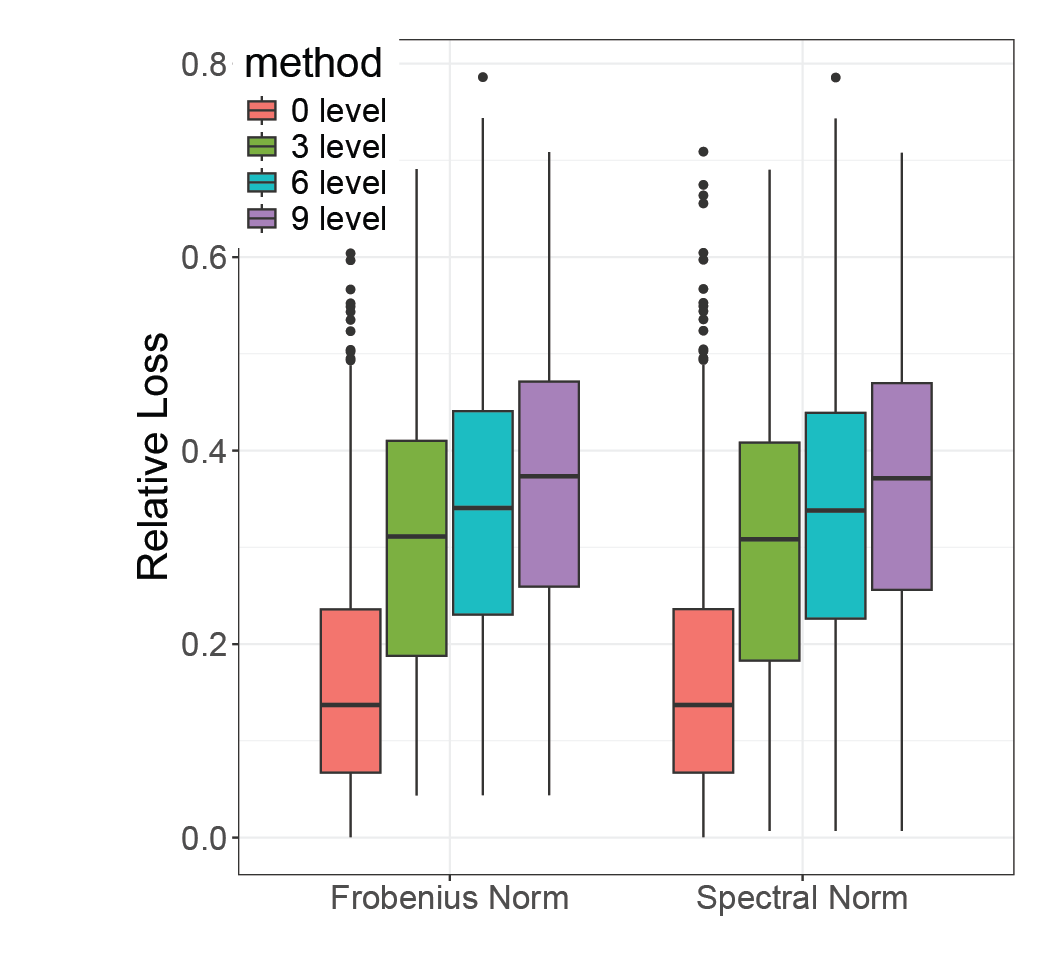} 
		\caption{
			We compare the relative losses of $\bm{\Sigma}_{\widehat{\bm{\beta}}}$ obtained from the MAUD under $n = 50$, $G = 3$, $\bm{\ell} = (30, 30, 40) ^ \top$, $R = 100$, and various noise levels,
			where ``0 level" denotes no noise, 
			``3 level", ``6 level", and ``9 level" denote the averages of random matrices $\textbf{E}_{\sigma}$ are $3 \textbf{I}_R$, $6 \textbf{I}_R$, and $9 \textbf{I}_R$, respectively.
		}
		\label{Fig:study_3}
	\end{figure}

	For each simulated dataset, we compute estimates for $\bm{\beta}$ and $\bm{\Sigma}_{\widehat{\bm{\beta}}}$. 
	Then, we similarly calculate the relative losses and present the resulting boxplots in Figure~\ref{Fig:study_3} based on all replicates. 
	We label the results as ``0 level”, ``3 level”, ``6 level”, and ``9 level” corresponding to the respective levels of $\sigma$.
	
	%%%%%%%%%%%%%%%%%%
	%%%% Figure 4 %%%%
	%%%%%%%%%%%%%%%%%%
	%%%%%%%%%%%%%%%%%%%%%
	%%%End of Figure 4 %%
	%%%%%%%%%%%%%%%%%%%%%
	
	As anticipated, the boxplots in Figure~\ref{Fig:study_3} illustrate increases in relative loss of $\bm{\Sigma}_{\widehat{\bm{\beta}}}$ as the noise level becomes larger. 
	We note that if $\sigma \neq 0$, then the average of random matrix $\textbf{E}_{\sigma}$ equals $\sigma R \textbf{I}_R \in \{3 \textbf{I}_R, 6 \textbf{I}_R, 9 \textbf{I}_R \}$.
	In other words, the elements of $\textbf{E}_{\sigma}$ approximate the same order of magnitude as those of the true covariance matrix. 
	The results in Figure~\ref{Fig:study_3} exhibit that the relative loss just doubles when the interconnected community structure is disrupted by noise of comparable magnitude.
	This suggests that the MAUD can effectively handle mild to moderate levels of noise in the covariance structure.
	
	\section{Investigation of the Effect of Alcohol Intake on Plasma Metabolomics}
	
	\label{Sec:real}
	
	To investigate the influence of alcohol consumption on plasma metabolomics, we applied our proposed model to a dataset from nuclear magnetic resonance (NMR) plasma metabolomics.
	Our primary focus was on understanding the associations between the NMR metabolic biomarkers and alcohol intake.
	The dataset is publicly accessible through the UK Biobank and has been comprehensively described by \citet{RitchieSurendranKarthikeyan2023}.
	It comprises $S = 249$ NMR metabolic biomarker measurements as outcomes.
	The exposure variable is alcohol intake frequency, with $912$ participants reporting daily intake, $1175$ three or four times a week, $1017$ once or twice a week, $424$ one to three times a month, $300$ on special occasions only, and $156$ reporting never consuming alcohol.
	Additionally, we considered the intercept, age (at the time of attending the assessment center, $63.39 \pm 7.61$ in year), sex (with $1840$ male and $2144$ female participants), BMI (body mass index, $26.28 \pm 4.11$), and heavy smoking (with $883$ participants categorized as smokers and $3101$ as non-smokers) as covariates. 
	In this dataset, we focused on $p = 6$ covariates with a total of $n = 3984$ participants.
	
	\textbf{\emph{Detecting the interconnected community structure in NMR data.}}
	We fitted the data to a multivariate linear regression model using the least-square method and obtained the estimated regression coefficient matrix $\widetilde{\mathfrak{B}}_{249 \times 6} ^ *$ of $\mathfrak{B}_{249 \times 6} ^ * = \left(\bm{\beta}_1 ^ *, \ldots, \bm{\beta}_{249} ^ * \right) ^ \top$. 
	We next calculated the residual matrix and applied a network detection algorithm proposed by \citet{ChenZhangWu2024}. 
	After reordering the biomarkers, the correlation matrix of the residuals revealed that $R = 170$ biomarkers are categorized into $G = 5$ interconnected communities with sizes of $L_1 = 77$, $L_2 = 47$, $L_3 = 19$, $L_4 = 11$, and $L_5 = 16$.
	The remaining $L_0 = 79$ biomarkers were found to be isolated (singletons), depicted in the heatmap in Figure~\ref{Fig:NMR} (B). 
	We also derived the estimated regression coefficient matrix $\widehat{\mathfrak{B}}_{249 \times 6} ^ *$ based on $\widetilde{\mathfrak{B}}_{249 \times 6} ^ *$, by arranging its rows according to the reordered biomarkers.
	In other words, the first $R = 170$ rows of $\widehat{\mathfrak{B}}_{249 \times 6} ^ *$ correspond to the estimates for the biomarkers in the interconnected communities, while the last $L_0 = 79$ rows represent the estimates for the singletons.
	The names of the biomarkers, their community indexes, their source communities as defined by \citet{RitchieSurendranKarthikeyan2023}, and $\widehat{\mathfrak{B}}_{249 \times 6} ^ *$, are available in the \href{Supplementary Material.pdf}{Supplementary Material}.

\begin{figure}[!h]
		\centering
		\includegraphics[width = 1.0 \linewidth]{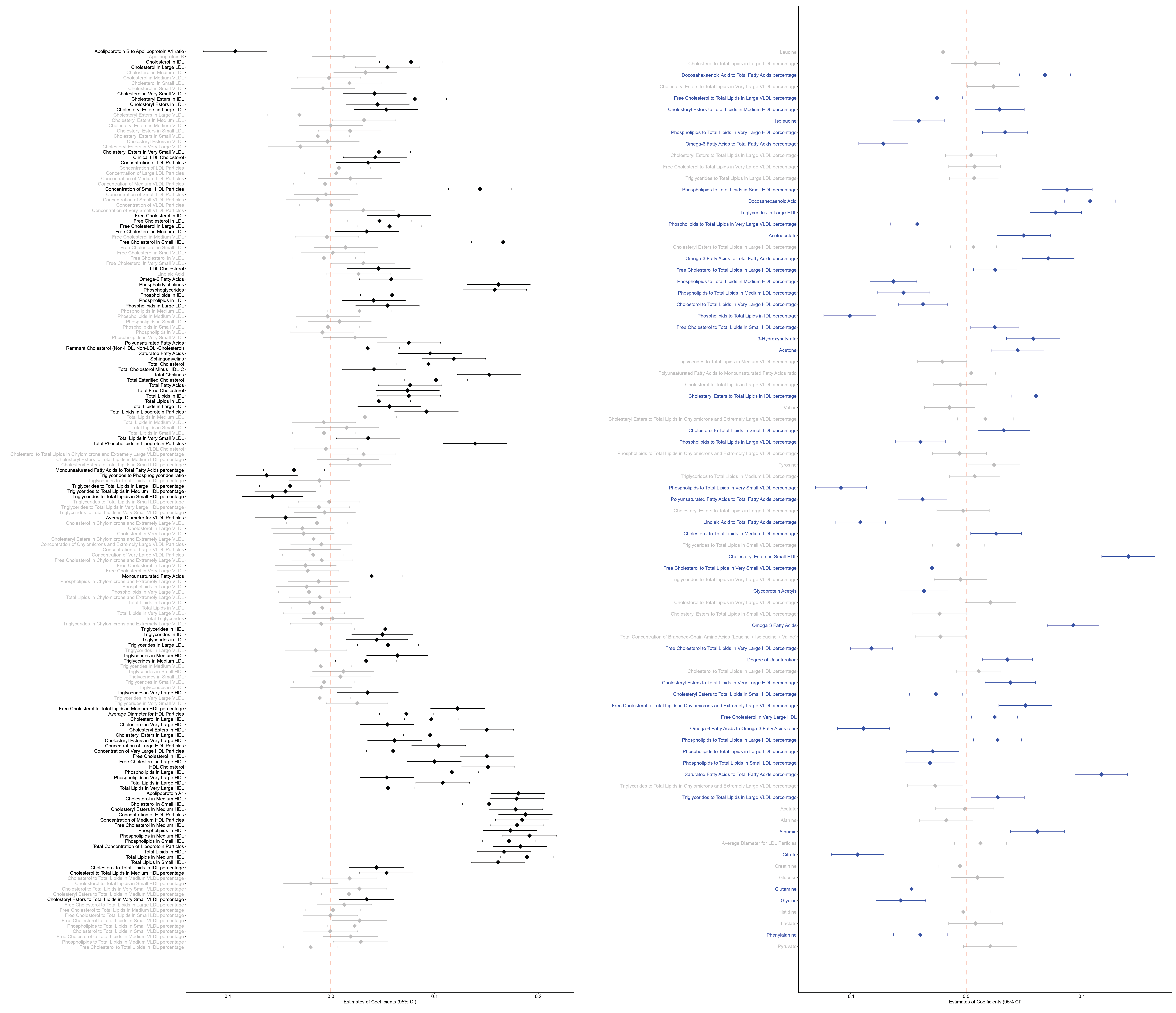} 
		\caption{
			We illustrate $95\%$ confidence intervals for the regression coefficients of $249$ biomarkers using a forest plot.
			Left: biomarker names in black correspond to biomarkers who are within the $5$ interconnected communities and their confidence intervals do not contain $0$; 
			Right: biomarker names in blue represent singleton biomarkers who are not in any communities and their confidence intervals do not contain $0$, 
			biomarker names in grey denote biomarkers whose confidence intervals contain $0$.
		}
		\label{Fig:real}
	\end{figure}

	\textbf{\emph{Statistical inference about $\mathfrak{B}_{249 \times 6} ^ *$.}}
	Utilizing the detected interconnected community structure, we aim to conduct statistical inference about the regression coefficient matrix $\mathfrak{B}_{249 \times 6} ^ *$ separately using the proposed MAUD and the general linear model.
	First, we partitioned $\widehat{\mathfrak{B}}_{249 \times 6} ^ *$ into two submatrices: one comprising its initial $170$ rows and the other containing the remaining $79$ rows.
	We then estimated the standard errors for the elements of the first submatrix using the proposed MAUD and for the elements of the second submatrix using the general linear model.
	The detailed results can be found in the \href{Supplementary Material.pdf}{Supplementary Material}.
	Subsequently, we proceeded to identify alcohol intake-related NMR biomarkers based on the inference outcomes at a false discovery rate (FDR) level of $0.05$.
	Additionally, we illustrated the $95\%$ confidence intervals for all regression coefficients in Figure~\ref{Fig:real}.

	%%%%%%%%%%%%%%%%%%
	%%%% Figure 5 %%%%
	%%%%%%%%%%%%%%%%%%
	%%%%%%%%%%%%%%%%%%%%%
	%%%End of Figure 5 %%
	%%%%%%%%%%%%%%%%%%%%%

	\textbf{\emph{Result.}}
	Our findings demonstrate significant positive associations with alcohol intake at a FDR level of $0.05$ for most high-density lipoprotein (HDL) biomarkers and the Apolipoprotein A1 biomarker.
	
	More specifically, within the pool of $249$ NMR biomarkers, $12$, $80$, $44$, and $58$ NMR biomarkers associated with the intermediate-density lipoprotein (IDL), very-low-density lipoprotein (VLDL), low-density lipoprotein (LDL), and HDL, respectively, displayed significant associations.
	A comprehensive result of the numbers of biomarkers within (or outside) the interconnected communities, along with their statistical decisions and the signs of estimated regression coefficients, is presented in Table~\ref{Tab:real}.

	%%%%%%%%%%%%%%%%%
	%%%% Table 2 %%%%
	% \usepackage{multirow}
	\begin{table}[!h]
		\centering
		\begin{tabular}{cccccccc}
			\hline
			& \multicolumn{3}{c}{Biomarkers in $5$ Communities} & \multicolumn{3}{c}{Singletons} & \multirow{3}{*}{Total} \\ \cline{2-7}
			& \multicolumn{2}{c}{significant} & \multirow{2}{*}{non-significant} & \multicolumn{2}{c}{significant} & \multirow{2}{*}{non-significant} &  \\ \cline{2-3} \cline{5-6}
			& $+$ & $-$ &  & $+$ & $-$ &  &  \\ \hline
			IDL & 8 & 0 & 2 & 1 & 1 & 0 & 12 \\
			VLDL & 4 & 1 & 56 & 2 & 5 & 12 & 80 \\
			LDL & 14 & 0 & 20 & 2 & 3 & 5 & 44 \\
			HDL & 36 & 3 & 3 & 10 & 4 & 2 & 58 \\ \hline
		\end{tabular}
		\caption{We provide a summary of the statistical decisions concerning $249$ NMR biomarkers, both within  and outside the interconnected communities, at a level of $0.05$,
			where ``IDL" denotes intermediate-density lipoprotein, 
			``VLDL" refers to very-low-density lipoprotein,
			``LDL" represents low-density lipoprotein, 
			``HDL" stands for high-density lipoprotein,
			and ``$+$" or ``$-$" signifies the sign of estimated regression coefficients, 
			indicating either a positive or negative association with alcohol intake.
		}
		\label{Tab:real}
	\end{table}
	%%%%%%%%%%%%%%%%%%%%%%%%
	%%%% End of Table 2 %%%%
	%%%%%%%%%%%%%%%%%%%%%%%%
	
	For LDL, the coefficients of $14$ significant NMR biomarkers among the first $170$ indicate a positive association, while the rest $5$ significant NMR biomarkers show both negative (i.e., phospholipids to total lipids in small/medium/large LDL percentage) and positive (i.e., cholesterol to total lipids in small/medium LDL percentage) associations. 
	
	In the case of HDL, $36$ NMR biomarkers out of the $39$ significant ones exhibit a positive association, while the remaining $3$ (i.e., triglycerides to total lipids in small/medium/large HDL percentage) are negatively associated with the alcohol intake frequency.
	Among the $14$ remaining significant biomarkers, alcohol intake frequency is negatively associated with  phospholipids to total lipids in medium HDL percentage, cholesterol to total lipids in very large HDL percentage, free cholesterol to total lipids in very large HDL percentage, and cholesteryl esters to total lipids in small HDL percentage only, with positive associations observed for the other biomarkers. 
	Summarily, out of $58$ HDL biomarkers, $36$ are positively associated with the frequency of alcohol intake, indicating that an increase in alcohol intake significantly affects these $36$ biomarkers in a positive manner. 
	In addition to the biomarkers linked to lipoprotein, we also observe a significant association between alcohol intake and Apolipoprotein B to Apolipoprotein A1 ratio.
	Additionally, the association between alcohol intake and Apolipoprotein A1 is significant positive, whereas the association with Apolipoprotein B is not significant at the level of $0.05$.
	This aligns with the notion that apolipoprotein A1 may likely play a pivotal role in raising HDL cholesterol levels as alcohol consumption increases \citep{SilvaFosterHarper2000}.
	
	In contrast, we also calculated the standard errors for $\widetilde{\mathfrak{B}}_{249 \times 6} ^ *$ and those for the first $R = 170$ rows of $\widehat{\mathfrak{B}}_{249 \times 6} ^ *$ using the GLMM. 
	Comprehensive inference results are available in the \href{Supplementary Material.pdf}{Supplementary Material}.

	\section{Discussion}
	
	\label{Sec:discussion}
	
	We have introduced a novel multivariate regression technique, the MAUD, designed to jointly analyze correlated multivariate outcomes. 
	In comparison to the conventional linear regression approaches that treat each outcome separately, the MAUD can effectively enhance statistical inference, i.e., it leads to higher statistical power or fewer false positive findings compared to competing methodologies. 
	This framework is broadly applicable in various omics data analyses, e.g., differential expression analysis, where datasets often exhibit underlying interconnected community structures.

	The MAUD is constructed using an autoregressive model while accommodating the interconnected community structure. 
	We bridge the gap between autoregressive dependence parameters and the parameters in a large covariance model. 
	By leveraging the interconnected community structure, we develop computationally efficient estimators for the dependence parameters in the autoregressive model.
	Accounting for the dependence among multivariate features in outcomes, we achieve more accurate inference, enabling precise selection of omics features.
	This advancement holds the potential to improve  reproducibility and replicability in high throughput data analysis.
	Notably, the MAUD exhibits robustness to model misspecification, as demonstrated in our simulation studies.
	While we utilized the interconnected community structure for the MAUD, a more comprehensive framework can be devised to incorporate a wider array of covariance structures under the autoregressive model.
	
	In our application to real-world data, we found evidence suggesting that alcohol intake can elevate the levels of ``good cholesterol" (HDL), potentially contributing to cardiovascular protection. 
	Furthermore, our observations regarding HDL and associated pathways involving Apolipoprotein A1 and Apolipoprotein B align well with known biological understanding.  
	However, caution is warranted regarding potential cancer risks associated with excessive alcohol consumption.  
	Collectively, our findings offer a new perspective on alcohol intake within the realm of public health.
	
	In conclusion, our proposed MAUD, along with its estimation and inference procedures, holds relevance and applicability in a wide spectrum of network studies characterized by an interconnected community structure.

	\newpage
	
	\bibliographystyle{biom}		
	
	\bibliography{paper_ref_MAUD}
	
\end{document}

\maketitle
	
	\begin{abstract}
		The present supplementary material contains four sections. 
		Section~\ref{Sec:ub_definitions} contains the definitions related to uniform-block matrix and uniform-block structure.
		Section~\ref{Sec:ub_properties} includes the properties of the uniform-block matrices, the closed-form transformations between uniform-block matrices $\textbf{N}_2 = \left(\textbf{I}_R - \textbf{N}_1\right)^{-1} \left(\textbf{I}_R - \textbf{N}_1 \right)^{-1}$, and the plug-in estimators of $\bm{\Upsilon}$, $\bm{\Omega}$, and $\bm{\Sigma}$.
		Section~\ref{Sec:conditions} details the technical conditions for main results in the manuscript. 
		Section~\ref{Sec:numerical} has the additional numerical results for both Simulation Studies and Investigation of the Effect of Alcohol Intake on Plasma Metabolomics sections. 
		Section~\ref{Sec:simulation} presents the additional simulation studies.
		Section~\ref{Sec:proof} provides the technical proofs and Section~\ref{Sec:case} considers the case of $\bm{\Sigma}_{\bm{\epsilon}} \neq \textbf{I}_{R}$.
		In addition to this document, the online material folder contains some additional simulation results presented in Excel files and the \proglang{R} code used for simulation studies in Section 3 and investigation of the effect of alcohol intake on plasma metabolomics in Section 4. 
		The code is also accessible at \url{https://github.com/yiorfun/MAUD}. 
	\end{abstract}
	
	\appendix
	
	\section{DEFINITIONS}
	
	\label{Sec:ub_definitions}
	
	\begin{dfns}[partition-size vector and partitioned matrix by a partition-size vector]
		\label{Dfn:par}
		Given positive integers $G < R$, 
		
		(1) if a column vector $\left(L_1, \ldots, L_G\right) ^ \top \in \mathbb{Z}_{+} ^ {G}$ satisfies that $R = L_1 + \cdots + L_G$ and $L_g > 1$ for all $g$, 
		then it is said to be a partition-size vector,  
		denoted by $\bm{\ell} = \left(L_1, \ldots, L_G\right) ^ \top$;
		
		(2) if a matrix $\textbf{N} \in \mathbb{R} ^ {R \times R}$ has a block form $\left(\textbf{N}_{gg'}\right)$ satisfying that the block $\textbf{N}_{gg'}$ has dimensions $L_g$ by $L_{g'}$ for $g, g' = 1, \ldots, G$,		
		then the partitioned matrix $\left(\textbf{N}_{gg'}\right)$ is said to be the partitioned matrix of $\textbf{N}$ by $\bm{\ell}$.
	\end{dfns}
	
	\begin{dfns}[uniform-block structure and uniform-block matrix]
		\label{Dfn:ub}
		Given a partition-size vector $\bm{\ell} = \left(L_1, \ldots, L_G\right) ^ \top$ satisfying that $R = L_1 + \cdots + L_G$ and $L_g > 1$ for all $g$, and a symmetric matrix $\textbf{N} \in \mathbb{R} ^ {R \times R}$, 
		let $\left(\textbf{N}_{gg'}\right)$ denote the partitioned matrix of $\textbf{N}$ by $\bm{\ell}$.
		If there exist $a_{gg}, b_{gg'} \in \mathbb{R}$ satisfying that the diagonal block $\textbf{N}_{gg} = a_{gg} \textbf{I}_{L_g} + b_{gg} \textbf{J}_{L_g}$ for $g = g'$, the off-diagonal block $\textbf{N}_{gg'} = b_{gg'} \textbf{1}_{L_g \times L_{g'}}$ with $b_{g'g} = b_{gg'}$ for $g \neq g'$, 		
		then the matrix structure is said to be a uniform-block (UB) structure, 
		and the partitioned matrix $\left(\textbf{N}_{gg'}\right)$ is said to be a uniform-block (UB) matrix,
		denoted by $\textbf{N}\left(\textbf{A}, \textbf{B}, \bm{\ell} \right)$,
		where $\textbf{A} = \operatorname{diag}\left(a_{11}, \dots, a_{GG}\right)$ is the $G$ by $G$ diagonal matrix and $\textbf{B} = \left(b_{gg'}\right)$ is the $G$ by $G$ symmetric matrix with $b_{gg'} = b_{g'g}$ for $g \neq g'$.
	\end{dfns}
	
	\begin{dfns}[consistent estimator of a covariance matrix, also see Definition 8.2.1 in \citet{FombyHillJohnson1984} or page 68 on \citet{Schmidt2020}]
		\label{Dfn:consistence}
		If a covariance matrix $\bm{\Sigma}$ depends on a finite number of parameters $\theta_1, \ldots, \theta_p$, 
		and if $\widehat{\bm{\Sigma}}$ depends on consistent estimators $\widehat{\theta}_1, \ldots, \widehat{\theta}_p$, then $\widehat{\bm{\Sigma}}$ is said to be a consistent estimator of $\bm{\Sigma}$.
	\end{dfns}
	
	\section{PROPERTIES OF UNIFORM-BLOCK MATRICES}
	
	\label{Sec:ub_properties}
	
	\begin{cors}[block Hadamard product representation of a uniform-block matrix]
		\label{Cor:ub_rep}
		Let $\bm{\ell} = \left(L_1, \dots, L_G\right) ^ \top$ be a partition-size vector satisfying that $R = L_1 + \cdots + L_G$ and $L_g > 1$ for all $g$, and $\textbf{N}\left(\textbf{A}, \textbf{B}, \bm{\ell} \right)$ be a uniform-block matrix with a diagonal matrix $\textbf{A} = \operatorname{diag}\left(a_{11}, \dots, a_{GG}\right)$ and a symmetric matrix $\textbf{B} = \left(b_{gg'}\right)$ with $b_{gg'} = b_{g'g}$ for $g \neq g'$.
		Then, 
		
		\begin{equation*}
			\textbf{N}\left(\textbf{A}, \textbf{B}, \bm{\ell} \right) 
			= \textbf{A} \circ \textbf{I}(\bm{\ell}) + \textbf{B} \circ \textbf{J}(\bm{\ell}),
		\end{equation*}
		holds uniquely for $\textbf{A}$ and $\textbf{B}$, where $\textbf{I}(\bm{\ell}) = \textbf{I} \left(\textbf{I}_G, \textbf{0}_{G \times G}, \bm{\ell} \right) = \operatorname{Bdiag}\left(\textbf{I}_{L_1}, \ldots, \textbf{I}_{L_G}\right)$ and $\textbf{J}(\bm{\ell}) = \textbf{J} \left(\textbf{0}_{G \times G}, \textbf{J}_{G}, \bm{\ell} \right) = \left(\textbf{1}_{L_g \times L_{g'}}\right)$,
		and $\circ$ denotes the block Hadamard product: $\textbf{A} \circ \textbf{I}(\bm{\ell})$ refers to the block-diagonal matrix $\operatorname{Bdiag}\left(a_{11} \textbf{I}_{L_1}, \ldots, a_{GG} \textbf{I}_{L_G}\right)$ and $\textbf{B} \circ \textbf{J}(\bm{\ell})$ refers to the symmetric matrix $\left(b_{gg'} \textbf{1}_{L_g \times L_{g'}}\right)$, respectively.
	\end{cors}
	
	\begin{cors}[operations of uniform-block matrices]
		
		\label{Cor:ops_all}
		
		Given uniform-block matrices $\textbf{N} = \textbf{N}\left(\textbf{A}, \textbf{B}, \bm{\ell} \right)$, $\textbf{N}_1 = \textbf{N}_1\left(\textbf{A}_{1}, \textbf{B}_{1}, \bm{\ell} \right)$ and $\textbf{N}_2 = \textbf{N}_2 \left(\textbf{A}_{2}, \textbf{B}_{2}, \bm{\ell} \right)$ with diagonal matrices $\textbf{A} = \operatorname{diag}\left(a_{11}, \dots, a_{GG}\right)$, $\textbf{A}_{1}$, $\textbf{A}_{2}$, symmetric matrices $\textbf{B} = \left(b_{gg'}\right)$, $\textbf{B}_{1}$, $\textbf{B}_{2}$, and a common partition-size vector $\bm{\ell} = \left(L_1, \ldots, L_G\right) ^ \top$ satisfying $L_g > 1$ for all $g$ and $R = L_1 + \cdots + L_G$, respectively,
		let $\bm{\Delta} = \textbf{A} + \textbf{B} \textbf{L}$, 
		where $\textbf{L} = \operatorname{diag}(L_1, \ldots, L_G)$. 
		
		(1) (Addition/Subtraction) suppose $\textbf{N} ^ * = \textbf{N}_1 \pm \textbf{N}_2$, then the partitioned matrix of $\textbf{N} ^ *$ by $\bm{\ell}$ is a uniform-block matrix, denoted by $\textbf{N} ^ * \left(\textbf{A}^*, \textbf{B}^*, \bm{\ell} \right)$, where $\textbf{A}^* = \textbf{A}_1 \pm \textbf{A}_2$ and $\textbf{B}^* = \textbf{B}_1 \pm \textbf{B}_2$;
		
		(2) (Square) suppose $\textbf{N}^* = \textbf{N} ^ 2$, then the partitioned matrix of $\textbf{N} ^ *$ by $\bm{\ell}$ is a uniform-block matrix, denoted by $\textbf{N} ^ *\left(\textbf{A}^*, \textbf{B}^*, \bm{\ell} \right)$, where $\textbf{A} ^ *  = \textbf{A} ^ 2$ and $\textbf{B}^* = \textbf{A} \textbf{B} + \textbf{B} \textbf{A} + \textbf{B} \textbf{L} \textbf{B}$;
		
		(3) (Eigenvalues) $\textbf{N}\left(\textbf{A}, \textbf{B}, \bm{\ell} \right)$ has $R$ eigenvalues in total, 
		those are $a_{gg}$ with multiplicity $(L_g - 1)$ for $g = 1, \dots, G$ and the rest $G$ eigenvalues are identical with those of $\bm{\Delta}$;
		
		(4) (Determinant) $\textbf{N}\left(\textbf{A}, \textbf{B}, \bm{\ell} \right)$ has the determinant $\left( \prod_{g = 1}^{G} a_{gg} ^ {L_g - 1} \right) \times \det\left( \bm{\Delta} \right)$; 
		
		(5) (Inverse) suppose $\textbf{N}$ is invertible and $\textbf{N}^* = \textbf{N} ^ {-1}$, then the partitioned matrix of $\textbf{N}^*$ by $\bm{\ell}$ is a UB matrix, denoted by $\textbf{N} ^ *\left(\textbf{A}^*, \textbf{B}^*, \bm{\ell} \right)$, where $\textbf{A}^* = \textbf{A} ^ {- 1}$ and $\textbf{B} ^ * = - \bm{\Delta} ^ {- 1} \textbf{B} \textbf{A} ^ {- 1}$.
		
		%(6) (Correlation Form) suppose $\textbf{N} ^ * = \operatorname{corr}\left(\textbf{N}\right)$ is the correlation matrix of $\textbf{N}$, then the partitioned matrix of $\textbf{N} ^ *$ by $\bm{r}$ is a UB matrix, denoted by $\textbf{N} ^ *\left(\textbf{A}^*, \textbf{B}^*, \bm{r} \right)$, where $\textbf{A} ^ * = \textbf{C} ^ {-1/2} \times \textbf{A} \times \textbf{C} ^ {-1/2}$, $\textbf{B} ^ * = \textbf{C} ^ {-1/2} \times \textbf{B} \times \textbf{C} ^ {-1/2}$, $\textbf{C} = \operatorname{diag} \left(c_{11}, \ldots, c_{GG}\right)$, and $c_{gg} = a_{gg} + b_{gg}$ for all $g$.
	\end{cors}
	
	\begin{cors}[square-inverse transformations]
		
		\label{Cor:sit}
		
		Let $\bm{\ell} = \left(L_1, \dots, L_G\right) ^ \top$ be a partition-size vector satisfying that $R = L_1 + \cdots + L_G$ and $L_g > 1$ for all $g$, and $\textbf{N}_1, \textbf{N}_2 \in \mathbb{R} ^ {R \times R}$ be two matrices satisfying the square-inverse relationship, i.e., $\textbf{N}_2 = \left(\textbf{I}_R - \textbf{N}_1 \right) ^ {- 1} \left(\textbf{I}_R - \textbf{N}_1\right) ^ {-1}$.
		
		(1) If $\textbf{N}_1 = \textbf{N}_1 \left(\textbf{A}_1, \textbf{B}_1, \bm{\ell} \right)$ is a UB matrix with diagonal matrix $\textbf{A}_1$ and symmetric matrix $\textbf{B}_1$, then $\textbf{N}_2$ is a UB matrix, expressed by $\textbf{N}_2 \left(\textbf{A}_2, \textbf{B}_2, \bm{\ell} \right)$.
		Specifically, let $\textbf{L} = \operatorname{diag}\left(L_1, \ldots, L_G\right)$,  
		$\textbf{A}_2$ and $\textbf{B}_2$ can be calculated using $\textbf{A}_1$, $\textbf{B}_1$, and $\textbf{L}$ as below:
		
		\begin{equation*}
			\begin{split}
				\textbf{I}_R - \textbf{N}_1 & = \textbf{N} ^ \ast \left(\textbf{A} ^ \ast, \textbf{B} ^ \ast, \bm{\ell} \right), 
				\quad \textbf{A} ^ \ast = \textbf{I}_G - \textbf{A}_1, 
				\quad \textbf{B} ^ \ast = - \textbf{B}_1, 
				\quad \bm{\Delta} ^ \ast = \textbf{I}_G - \textbf{A}_1 - \textbf{B}_1 \textbf{L}; \\
				\left(\textbf{I}_R - \textbf{N}_1\right) ^ {-1} & = \textbf{N} ^ \star \left(\textbf{A} ^ \star, \textbf{B} ^ \star, \bm{\ell} \right), 
				\quad \textbf{A} ^ \star = \textbf{A} ^ {\ast, - 1},
				\quad \textbf{B} ^ \star = - \bm{\Delta} ^ {\ast, - 1} \textbf{B} ^ \ast \textbf{A} ^ {\ast, - 1}; \\
				\left(\textbf{I}_R - \textbf{N}_1 \right) ^ {-1} \left(\textbf{I}_R - \textbf{N}_1\right) ^ {-1}
				& = \textbf{N}_2 \left(\textbf{A}_2, \textbf{B}_2, \bm{\ell} \right), 
				\quad \textbf{A}_2 = \textbf{A} ^ {\star, 2}, 
				\quad \textbf{B}_2 = \textbf{A} ^ \star \textbf{B} ^ \star + \textbf{B} ^ \star \textbf{A} ^ \star + \textbf{B} ^ \star \textbf{L} \textbf{B} ^ \star.
			\end{split}
		\end{equation*}
		
		(2) If $\textbf{N}_2 = \textbf{N}_2\left(\textbf{A}_2, \textbf{B}_2, \bm{\ell} \right)$ is a UB matrix with diagonal matrix $\textbf{A}_2$ and symmetric matrix $\textbf{B}_2$, then $\textbf{N}_1$ is a UB matrix, expressed by $\textbf{N}_1\left(\textbf{A}_1, \textbf{B}_1, \bm{\ell} \right)$.
		Specifically, let $\textbf{L} = \operatorname{diag}\left(L_1, \ldots, L_G\right)$,  
		$\textbf{A}_1$ and $\textbf{B}_1$ can be calculated using $\textbf{A}_2$, $\textbf{B}_2$, $\textbf{L}$, and $\bm{\Delta}_2 = \textbf{A}_2 + \textbf{B}_2 \textbf{L}$ as below:
		
		\begin{equation*}
			\begin{split}
				\textbf{N}_2 ^ {-1} & = \textbf{N} ^ \dagger \left(\textbf{A} ^ \dagger, \textbf{B} ^ \dagger, \bm{\ell} \right), 
				\quad \textbf{A} ^ \dagger = \textbf{A}_2 ^ {- 1}, 
				\quad \textbf{B} ^ \dagger = - \bm{\Delta}_2 ^ {- 1} \textbf{B}_2 \textbf{A}_2 ^ {-1}; \\
				\textbf{I}_R - \textbf{N}_1 & = \textbf{N} ^ \ast \left(\textbf{A} ^ \ast, \textbf{B} ^ \ast, \bm{\ell} \right),
				\quad \textbf{A} ^ \ast = \textbf{A} ^ {\dagger, \frac{1}{2}}, 
				\quad \textbf{B} ^ \ast \text{ is the solution to the algebraic Riccati equation: } \\
				& \textbf{A} ^ \ast \times \textbf{B} ^ \ast + \textbf{B} ^ \ast \times \textbf{A} ^ \ast + \textbf{B} ^ \ast \times \textbf{L} \times \textbf{B} ^ \ast - \textbf{B} ^ {\dagger} = \textbf{0}_{G \times G}; \\
				\textbf{N}_1 & = \textbf{N}_1 \left(\textbf{A}_1, \textbf{B}_1, \bm{\ell}\right), 
				\quad \textbf{A}_1 = \textbf{I}_G - \textbf{A} ^ \ast, 
				\quad \textbf{B}_1 = - \textbf{B} ^ \ast.
			\end{split}
		\end{equation*}
	\end{cors}
	
	\begin{cors}[plug-in estimators of matrix parameters]
		
		\label{Cor:estimators}
		
		The plug-in matrix estimators are given by
		\begin{align}
			\label{Eq:Upsilon_AB_EST}
			\widehat{\bm{\Upsilon}}\left(\widehat{\textbf{A}}_{\bm{\Upsilon}}, \widehat{\textbf{B}}_{\bm{\Upsilon}}, \bm{\ell} \right) & = \widehat{\textbf{A}}_{\bm{\Upsilon}} \circ \textbf{I}(\bm{\ell}) + \widehat{\textbf{B}}_{\bm{\Upsilon}} \circ \textbf{J}(\bm{\ell}),
			\text{ with } 
			\begin{cases}
				\widehat{\textbf{A}}_{\bm{\Upsilon}} & = \operatorname{diag}\left(- \widehat{\gamma}_{11}, \ldots, - \widehat{\gamma}_{GG}\right) \\
				\widehat{\textbf{B}}_{\bm{\Upsilon}} & = \left(\widehat{\gamma}_{gg'}\right)
			\end{cases}, \\
			%%%
			\label{Eq:Omega_AB_EST}
			\widehat{\bm{\Omega}}\left(\widehat{\textbf{A}}_{\bm{\Omega}}, \widehat{\textbf{B}}_{\bm{\Omega}}, \bm{\ell} \right) & = \widehat{\textbf{A}}_{\bm{\Omega}} \circ \textbf{I}(\bm{\ell}) + \widehat{\textbf{B}}_{\bm{\Omega}} \circ \textbf{J}(\bm{\ell}), 
			\text{ with } 
			\begin{cases}
				\widehat{\textbf{A}}_{\bm{\Omega}} & = \left(\textbf{I}_{G} - \widehat{\textbf{A}}_{\bm{\Upsilon}} \right) ^ 2 \\
				\widehat{\textbf{B}}_{\bm{\Omega}} & = - 2 \widehat{\textbf{B}}_{\bm{\Upsilon}} + \widehat{\textbf{A}}_{\bm{\Upsilon}}   \widehat{\textbf{B}}_{\bm{\Upsilon}} + \widehat{\textbf{B}}_{\bm{\Upsilon}}  \widehat{\textbf{A}}_{\bm{\Upsilon}} + \widehat{\textbf{B}}_{\bm{\Upsilon}}  \textbf{L}  \widehat{\textbf{B}}_{\bm{\Upsilon}}
			\end{cases}, \\
			%%%
			\label{Eq:Sigma_AB_EST}
			\widehat{\bm{\Sigma}}\left(\widehat{\textbf{A}}_{\bm{\Sigma}}, \widehat{\textbf{B}}_{\bm{\Sigma}}, \bm{\ell} \right) & = \widehat{\textbf{A}}_{\bm{\Sigma}} \circ \textbf{I}(\bm{\ell}) + \widehat{\textbf{B}}_{\bm{\Sigma}} \circ \textbf{J}(\bm{\ell}),
			\text{ with } 
			\begin{cases}
				\widehat{\textbf{A}}_{\bm{\Sigma}} & = \widehat{\textbf{A}}_{\bm{\Omega}} ^ {- 1} \\
				\widehat{\textbf{B}}_{\bm{\Sigma}} & = - \widehat{\bm{\Delta}}_{\bm{\Omega}} ^ {- 1}  \widehat{\textbf{B}}_{\bm{\Omega}}  \widehat{\textbf{A}}_{\bm{\Omega}} ^ {- 1}
			\end{cases},
			%\widehat{\textbf{A}}_{\bm{\Xi}} & = \widehat{\textbf{C}}_{\bm{\Sigma}} ^ {- 1/2}  \widehat{\textbf{A}}_{\bm{\Sigma}}  \widehat{\textbf{C}}_{\bm{\Sigma}} ^ {- 1/2}, \quad
			%&& \widehat{\textbf{B}}_{\bm{\Xi}} = \widehat{\textbf{C}}_{\bm{\Sigma}} ^ {- 1/2}  \widehat{\textbf{B}}_{\bm{\Sigma}}  \widehat{\textbf{C}}_{\bm{\Sigma}} ^ {- 1/2},
			%\label{Eq:Xi_A_B_EST}
		\end{align}
		where we assume $\widehat{\textbf{A}}_{\bm{\Omega}} \succ 0$, $\widehat{\bm{\Delta}}_{\bm{\Omega}} =  \widehat{\textbf{A}}_{\bm{\Omega}} + \widehat{\textbf{B}}_{\bm{\Omega}} \textbf{L}$ has positive eigenvalues only.
		
		By Theorem 2, 
		we note that the matrix estimators $\widehat{\bm{\Upsilon}} \left(\widehat{\textbf{A}}_{\bm{\Upsilon}}, \widehat{\textbf{B}}_{\bm{\Upsilon}}, \bm{\ell}\right)$ in~\eqref{Eq:Upsilon_AB_EST}, $\widehat{\bm{\Omega}} \left(\widehat{\textbf{A}}_{\bm{\Omega}}, \widehat{\textbf{B}}_{\bm{\Omega}}, \bm{\ell}\right)$ in~\eqref{Eq:Omega_AB_EST}, and $\widehat{\bm{\Sigma}} \left(\widehat{\textbf{A}}_{\bm{\Sigma}}, \widehat{\textbf{B}}_{\bm{\Sigma}}, \bm{\ell}\right)$ in~\eqref{Eq:Sigma_AB_EST} %and $\widehat{\bm{\Xi}} \left(\widehat{\textbf{A}}_{\bm{\Xi}}, \widehat{\textbf{B}}_{\bm{\Xi}}, \bm{r}\right)$ in~\eqref{Eq:Xi_EST} 
		are consistent estimators in the sense of Definition~\ref{Dfn:consistence} in Section~\ref{Sec:ub_definitions}.
	\end{cors}
	
	\section{TECHNICAL CONDITIONS}
	
	\label{Sec:conditions}
	
	\begin{condition}
		\label{Con:pd}
		Covariance matrix $\bm{\Sigma}\left(\textbf{A}_{\bm{\Sigma}}, \textbf{B}_{\bm{\Sigma}}, \bm{\ell}\right)$ and the proposed estimator $\widehat{\bm{\Sigma}} \left(\widehat{\textbf{A}}_{\bm{\Sigma}}, \widehat{\textbf{B}}_{\bm{\Sigma}}, \bm{\ell}\right)$ are positive definite, or equivalently, $\textbf{A}_{\bm{\Sigma}}, \widehat{\textbf{A}}_{\bm{\Sigma}} \succ 0$ and both $\bm{\Delta}_{\bm{\Sigma}} = \textbf{A}_{\bm{\Sigma}} + \textbf{B}_{\bm{\Sigma}} \textbf{L}$ and $\widehat{\bm{\Delta}}_{\bm{\Sigma}} = \widehat{\textbf{A}}_{\bm{\Sigma}} + \widehat{\textbf{B}}_{\bm{\Sigma}} \textbf{L}$ have positive eigenvalues only.
	\end{condition}
	
	\begin{condition}
		\label{Con:linear_independent}
		Covariate vectors $\textbf{x}_1, \ldots, \textbf{x}_n \in \mathbb{R} ^ {p}$ are linear independent.
	\end{condition}
	
	\begin{condition}
		\label{Con:sample_size}
		Sample size $n > \max\left\{p, G (G + 1) / 2\right\}$.
	\end{condition}
	
	\begin{condition}
		\label{Con:beta_gamma}
		Unknown regression coefficient vector $\bm{\beta}$ does not depend on unknown scaled dependence parameter vector $\bm{\gamma}$.
	\end{condition}
	
	\begin{condition}
		\label{Con:beta_converge}
		Each element of $n ^ {-1}\bm{\Phi}_{\bm{\beta}}$ converges to a finite function with respect to $\bm{\gamma}$, as $n$ goes to infinity, uniformly for $\bm{\gamma}$ in the compact set $\Theta$.
	\end{condition}
	
	\begin{condition}
		\label{Con:diagonal_converge}
		Each diagonal element of 
		\begin{equation*}
			n ^ {- 2} \left\{ \left[\dfrac{\partial \bm{\Omega}\left(\textbf{A}_{\bm{\Omega}}, \textbf{B}_{\bm{\Omega}}, \bm{\ell}\right)}{\partial \bm{\gamma}_j} \right] \bm{\Sigma} \left(\textbf{A}_{\bm{\Sigma}}, \textbf{B}_{\bm{\Sigma}}, \bm{\ell} \right) \left[\dfrac{\partial \bm{\Omega}\left(\textbf{A}_{\bm{\Omega}}, \textbf{B}_{\bm{\Omega}}, \bm{\ell}\right)}{\partial \bm{\gamma}_j} \right]\right \} \otimes \left(\sum_{i = 1}^{n} \textbf{x}_i \textbf{x}_i ^ \top \right)
		\end{equation*}
		converges to $0$, as $n$ goes to infinity, uniformly for $\bm{\gamma}$ in the compact set $\Theta$, for all $j$.
	\end{condition}
	
	\begin{section}{ADDITIONAL NUMERICAL RESULTS}
		
		\label{Sec:numerical}
		
		\subsection{Simulation studies}
		
		\label{Subsec:FDR}
		
		We conducted an assessment of the statistical inference regarding the regression coefficients using the MAUD and compare it with approaches of AMBD, MMRM, and GLMM. 
		As a standard, we also utilized the true covariance matrix $\bm{\Sigma}\left(\bm{\textbf{A}}_{\bm{\Sigma}}, \textbf{B}_{\bm{\Sigma}}, \bm{\ell} \right)$ as an ``estimator”, denoted by ``TRUE”.
		Our primary focus is on identifying a subset of features associated with the covariates. 
		In addition to calculation the relative loss of $\bm{\Sigma}_{\widehat{\bm{\beta}}}$, we perform the hypotheses below:
		\begin{equation*}
			H_{0, r, 2}: \beta_r ^ {(2)} = 0 \quad \text{versus}
			H_{1, r, 2}: \beta_r ^ {(2)} \neq 0, \quad r = 1, \ldots, R.
		\end{equation*}
		
		Specifically, for each replication, we estimated both $\bm{\beta}$ $\bm{\Sigma}_{\widehat{\bm{\beta}}}$ using the MAUD, AMBD, MMRM, and GLMM, respectively.
		We then calculated the $t$-value $\widehat{\beta}_r ^ {(2)} / \operatorname{SE}\left(\widehat{\beta}_r ^ {(2)}\right)$ with a degree of freedom $(n - 1)$ for each $r$, where $\operatorname{SE}\left(\widehat{\beta}_r ^ {(2)}\right)$ could be obtained from the diagonal entries of $\widehat{\Sigma}_{\widehat{\bm{\beta}}}$.
		Next, since these $t$-values are almost independent (the off-diagonal elements of $\widehat{\Sigma}_{\widehat{\bm{\beta}}}$ were extremely small), 
		we calculated the adjusted $p$-values using the Benjamini-Hochberg procedure \citep{BenjaminiHochberg1995}, controlling the false discovery rate (FDR) at level of $\alpha = 0.05$.  
		Lastly, we reject the hypothesis if the corresponding adjusted $p$-value is smaller than $\alpha$.
		Among all replications, we calculated the empirical proportion of rejections for each $\beta_r ^ {(2)}$, 
		which indicated the type $1$ error if the truth $\beta_r ^ {(2)} = 0$ and indicated the statistical power if the truth $\beta_r ^ {(2)} \neq 0$.
		The results are provided in an Excel file available in the online material,
		demonstrating the MAUD exhibited the closest behaviors to the TRUE. 
		
	\end{section}
	
	\begin{section}{ADDITIONAL SIMULATION STUDIES}
		
		\label{Sec:simulation}
		
	\end{section}

	\section{PROOFS}
	
	\label{Sec:proof}
	
	\begin{proof}[Proofs of Corollary~\ref{Cor:ub_rep}, Corollary~\ref{Cor:ops_all}, and Corollary~\ref{Cor:estimators}]
		
		Please refer to the arguments and proofs in \citet{YangChenChen2024}.
	\end{proof}
	
	\begin{proof}[Proof of Corollary~\ref{Cor:sit}]
		Using the results (1), (2), and (5) in Corollary~\ref{Cor:ops_all}, 
		the proof is straightforward.		
		The requirement for a unique solution to the algebraic Riccati equation can be found in \citet{RanRodman1984} and \citet{AbouKandilFreilingIonescu2003}.
	\end{proof}
	
	\begin{proof}[Proof of Corollay 1] %~\ref{Cor:SigmaUB}]
		Using the result in Corollary~\ref{Cor:sit} and the result (6) in Corollary B.2, %~\ref{Cor:ops_all}, 
		the proof is straightforward.
	\end{proof}
	
	\begin{proof}[Proof of Theorem 1] %~\ref{Thm:estimator_properties}]
		
		To show (1), we check that Conditions~\ref{Con:pd},~\ref{Con:linear_independent},~\ref{Con:sample_size},~\ref{Con:beta_gamma} satisfy Assumptions 1-4 in \citet{Magnus1978}.
		In particular, Conditions~\ref{Con:linear_independent} and~\ref{Con:sample_size} imply the design matrix $\textbf{x} \in \mathbb{R} ^ {(nR) \times (Rp)}$ has a full rank; and the partial derivatives~\eqref{Eq:partial1} and~\eqref{Eq:partial2} imply that each element of the matrix $\bm{\Sigma}\left(\textbf{A}_{\bm{\Sigma}}, \textbf{B}_{\bm{\Sigma}}, \bm{\ell}\right)$ are twice differentiable function with respect to $\bm{\gamma}_j$, 
		where $\bm{\gamma}_j$ belonging to $\{\gamma_{11}, \ldots, \gamma_{1G}, \ldots, \gamma_{2G}, \ldots, \gamma_{GG}\}$ denotes the $j$-th component of $\bm{\gamma}$ for $j = 1, \ldots, G(G + 1) / 2$.
		Therefore, (1) holds by Theorem 1 in \citet{Magnus1978}.
		
		To prove (2), we start with the following equality
		\begin{equation*}
			\textbf{x}_{(Rp) \times (nR)} ^ \top \left(\textbf{I}_{n} \otimes \textbf{N}_{R \times R} \right)_{(nR) \times (nR)} \textbf{x}_{(nR) \times (Rp)}
			= \textbf{N} \otimes \left(\sum_{i = 1}^{n} \textbf{x}_i \textbf{x}_i ^ \top \right)
		\end{equation*}
		for $\textbf{N} \in \mathbb{R} ^ {R \times R}$.
		Taking $\textbf{N} = \textbf{I}_{R}$, we obtain that an alternative form of $\textbf{x} ^ \top \textbf{x}$, i.e., $\textbf{x} ^ \top \textbf{x} = \textbf{I}_{R} \otimes \left(\sum_{i = 1}^{n} \textbf{x}_i \textbf{x}_i ^ \top \right)$.
		By Condition~\ref{Con:linear_independent}, $\left(\sum_{i = 1}^{n} \textbf{x}_i \textbf{x}_i ^ \top \right)$ is invertible, so is $\textbf{x}^\top \textbf{x}$.
		Suppose $\textbf{N}$ is invertible, then 
		\begin{equation*}
			\begin{split}
				\left(\textbf{x} ^ \top \textbf{x}\right) ^ {- 1} & \textbf{x} ^ \top \left(\textbf{I}_{n} \otimes \textbf{N}\right) \textbf{x} \left(\textbf{x} ^ \top \textbf{x}\right) ^ {- 1} 
				= \left[\textbf{I}_R \otimes \left(\sum_{i = 1}^{n} \textbf{x}_i \textbf{x}_i ^ \top \right) ^ {- 1} \right]
				\left[\textbf{N} \otimes \left(\sum_{i = 1}^{n} \textbf{x}_i \textbf{x}_i ^ \top \right) \right]
				\left[\textbf{I}_R \otimes \left(\sum_{i = 1}^{n} \textbf{x}_i \textbf{x}_i ^ \top \right) ^ {- 1} \right] \\
				& = \textbf{N} \otimes \left(\sum_{i = 1}^{n} \textbf{x}_i \textbf{x}_i ^ \top \right) ^ {- 1} 
				= \left[\textbf{N} ^ {- 1}  \otimes \left(\sum_{i = 1}^{n} \textbf{x}_i \textbf{x}_i ^ \top \right) \right] ^ { - 1} 
				= \left[\textbf{x} ^ \top \left(\textbf{I}_{n} \otimes \textbf{N} \right) ^ {- 1} \textbf{x} \right] ^ {- 1}.
			\end{split}
		\end{equation*} 
		Following Condition~\ref{Con:pd} and Theorem 1(A) in \citet{LuSchmidt2012}, 
		we replace $\textbf{N} = \bm{\Sigma}\left(\textbf{A}_{\bm{\Sigma}}, \textbf{B}_{\bm{\Sigma}}, \bm{\ell}\right)$ and $\textbf{N} = \widehat{\bm{\Sigma}}\left(\widehat{\textbf{A}}_{\bm{\Sigma}}, \widehat{\textbf{B}}_{\bm{\Sigma}}, \bm{\ell} \right)$, respectively, and complete the proof of the equality of the OLS estimator and the FGLS estimator (see more discussions in \citet{PuntanenStyan1989} and a case of $G = 1$ in \citet{HeWang2022}).
		Under the normality assumption for the MAUD, %~\eqref{Eq:maud_vec}, 
		we obtain the normal distribution for $\widehat{\bm{\beta}}$.
		
		The proof of (3) follows the lines of arguments in the proof of Theorem 3 in \citet{Magnus1978}.
		Specifically, 
		let $\bm{\gamma}_j \in \{\gamma_{11}, \ldots, \gamma_{1G}, \ldots, \gamma_{2G}, \ldots, \gamma_{GG}\}$ denote the $j$-th component of $\bm{\gamma}$ for $j = 1, \ldots, G(G + 1) / 2$. 
		Then, $\bm{\Phi}_{\bm{\gamma}} = \left(\psi_{jj'} ^ {(\bm{\gamma})}\right)$ and $\psi_{jj'} ^ {(\bm{\gamma})}$ is given by  
		\begin{equation*}
			\begin{split}
				\psi_{jj'} ^ {(\bm{\gamma})}
				& = \frac{1}{2} \operatorname{tr} \bigg \{ \left[\frac{\partial \left(\textbf{I}_n \otimes \bm{\Sigma} \left(\textbf{A}_{\bm{\Sigma}}, \textbf{B}_{\bm{\Sigma}}, \bm{\ell}\right)\right) ^ {- 1}}{\partial \bm{\gamma}_j}\right] \\
				& \times
				\left(\textbf{I}_n \otimes \bm{\Sigma} \left(\textbf{A}_{\bm{\Sigma}}, \textbf{B}_{\bm{\Sigma}}, \bm{\ell}\right)\right) 
				\left[\frac{\partial \left(\textbf{I}_n \otimes \bm{\Sigma} \left(\textbf{A}_{\bm{\Sigma}}, \textbf{B}_{\bm{\Sigma}}, \bm{\ell}\right)\right) ^ {- 1}}{\partial \bm{\gamma}_{j'}}\right]
				\left(\textbf{I}_n \otimes \bm{\Sigma} \left(\textbf{A}_{\bm{\Sigma}}, \textbf{B}_{\bm{\Sigma}}, \bm{\ell}\right)\right) 
				\bigg \} \\
				& = \frac{n}{2} \operatorname{tr} \left\{
				\left[\frac{\partial \bm{\Omega}\left(\textbf{A}_{\bm{\Omega}}, \textbf{B}_{\bm{\Omega}}, \bm{\ell}\right) }{\partial \bm{\gamma}_j}\right] \bm{\Sigma} \left(\textbf{A}_{\bm{\Sigma}}, \textbf{B}_{\bm{\Sigma}}, \bm{\ell}\right)
				\left[\frac{\partial \bm{\Omega} \left(\textbf{A}_{\bm{\Omega}}, \textbf{B}_{\bm{\Omega}}, \bm{\ell}\right)}{\partial \bm{\gamma}_{j'}}\right] \bm{\Sigma} \left(\textbf{A}_{\bm{\Sigma}}, \textbf{B}_{\bm{\Sigma}}, \bm{\ell}\right)
				\right\}
			\end{split}
		\end{equation*}
		for $j, j' = 1, \ldots, G(G + 1) / 2$, 
		where $\dfrac{\partial \bm{\Omega} \left(\textbf{A}_{\bm{\Omega}}, \textbf{B}_{\bm{\Omega}}, \bm{\ell}\right) }{\partial \bm{\gamma}_j}
		= \left(\dfrac{\partial \textbf{A}_{\bm{\Omega}}}{\partial \bm{\gamma}_j}\right) \circ \textbf{I}(\bm{\ell})
		+ \left(\dfrac{\partial \textbf{B}_{\bm{\Omega}}}{\partial \bm{\gamma}_j}\right) \circ \textbf{J}(\bm{\ell})$.
		Recall the following ``coordinate matrices", 
		\begin{alignat*}{3}
			\textbf{A}_{\bm{\Upsilon}} & = \operatorname{diag}\left(- \gamma_{11}, \ldots, - \gamma_{GG} \right), \quad
			&& \textbf{B}_{\bm{\Upsilon}} && = \left(\gamma_{gg'}\right); \\
			\textbf{A}_{\bm{\Omega}} & = \left(\textbf{I}_{G} - \textbf{A}_{\bm{\Upsilon}} \right) ^ 2, \quad 
			&& \textbf{B}_{\bm{\Omega}} && = - 2 \textbf{B}_{\bm{\Upsilon}} + \textbf{A}_{\bm{\Upsilon}}   \textbf{B}_{\bm{\Upsilon}} + \textbf{B}_{\bm{\Upsilon}}  \textbf{A}_{\bm{\Upsilon}} + \textbf{B}_{\bm{\Upsilon}}  \textbf{L}  \textbf{B}_{\bm{\Upsilon}};  \\
			\textbf{A}_{\bm{\Sigma}} & = \textbf{A}_{\bm{\Omega}} ^ {- 1}, \quad
			&& \textbf{B}_{\bm{\Sigma}} && = - \bm{\Delta}_{\bm{\Omega}} ^ {- 1}  \textbf{B}_{\bm{\Omega}}  \textbf{A}_{\bm{\Omega}} ^ {- 1}; 
		\end{alignat*}
		and let $\textbf{E}_{gg'} \in \mathbb{R} ^ {G \times G}$ denote a matrix whose $(g,g')$-th element is $1$ and the others are $0$.
		
		If $\bm{\gamma}_j = \gamma_{gg}$ for some $g$, then
		$\dfrac{\partial \bm{\Omega} \left(\textbf{A}_{\bm{\Omega}}, \textbf{B}_{\bm{\Omega}}, \bm{\ell}\right)}{\partial \bm{\gamma}_j}
		= \left(\dfrac{\partial \textbf{A}_{\bm{\Omega}}}{\partial \bm{\gamma}_j}\right) \circ \textbf{I}(\bm{\ell})
		+ \left(\dfrac{\partial \textbf{B}_{\bm{\Omega}}}{\partial \bm{\gamma}_j}\right) \circ \textbf{J}(\bm{\ell})$, 
		and 
		\begin{equation}
			\begin{split}
				\label{Eq:partial1}
				\frac{\partial \textbf{A}_{\bm{\Omega}}}{\partial \bm{\gamma}_j}
				& = 2 (1 + \gamma_{gg}) \textbf{E}_{gg}, \\
				%%%
				\frac{\partial \textbf{B}_{\bm{\Omega}}}{\partial \bm{\gamma}_j}
				& = - 2(1 + \gamma_{gg}) \textbf{E}_{gg} 
				- \left(\textbf{E}_{gg} \textbf{B}_{\bm{\Upsilon}} 
				%- \gamma_{gg} \textbf{E}_{gg}
				%- \gamma_{gg} \textbf{E}_{gg}
				+ \textbf{B}_{\bm{\Upsilon}} \textbf{E}_{gg} \right)
				+ \textbf{E}_{gg} \textbf{L} \textbf{B}_{\bm{\Upsilon}} 
				+ \textbf{B}_{\bm{\Upsilon}} \textbf{L} \textbf{E}_{gg}.
			\end{split}
		\end{equation}
		If $\bm{\gamma}_{j} = \gamma_{gg'}$ for some $g \neq g'$, then 
		$\dfrac{\partial \bm{\Omega} \left(\textbf{A}_{\bm{\Omega}}, \textbf{B}_{\bm{\Omega}}, \bm{\ell}\right)}{\partial \bm{\gamma}_j}
		= \textbf{0}_{G \times G} \circ \textbf{I}(\bm{\ell})
		+ \left(\dfrac{\partial \textbf{B}_{\bm{\Omega}}}{\partial \bm{\gamma}_j}\right) \circ \textbf{J}(\bm{\ell})$, 
		and 
		\begin{equation}
			\begin{split}
				\label{Eq:partial2}
				\frac{\partial \textbf{A}_{\bm{\Omega}}}{\partial \bm{\gamma}_j}
				& = \textbf{0}_{G \times G}, \\
				%%%
				\frac{\partial \textbf{B}_{\bm{\Omega}}}{\partial \bm{\gamma}_j}
				& = 
				- 2 \left(\textbf{E}_{gg'} + \textbf{E}_{g'g}\right)
				+ \left(\textbf{E}_{gg'} + \textbf{E}_{g'g}\right) \left(\textbf{A}_{\bm{\Upsilon}} + \textbf{L} \textbf{B}_{\bm{\Upsilon}} \right)
				+ \left(\textbf{A}_{\bm{\Upsilon}} + \textbf{B}_{\bm{\Upsilon}} \textbf{L} \right) \left(\textbf{E}_{gg'} + \textbf{E}_{g'g}\right).
			\end{split}
		\end{equation}
		Due to $\textbf{E}_{gg'}$ for $g, g' = 1, \ldots, G$, 
		$\operatorname{vec}\left[\partial \bm{\Omega}\left(\textbf{A}_{\bm{\Omega}}, \textbf{B}_{\bm{\Omega}}, \bm{\ell}\right) / \partial \bm{\gamma}_j\right]$ for $j = 1, \ldots, G(G + 1) / 2$ are linearly independent, 
		yielding Assumption 5 in \citet{Magnus1978} is satisfied.
		Thus, $\bm{\Psi} \succ 0$ follows the result of Lemma 1 in \citet{Magnus1978}.
		
		The proof of (4) follows the lines of the proof of Theorem 5 in \citet{Magnus1978}.
		Given the first-order partial derivatives in closed-form, 
		we can obtain the second-order partial derivatives in closed-form, which is omitted here. 
		So, Assumption 10 and Assumption 11 in \citet{Magnus1978} are satisfied because both the following facts:
		$2 n ^ {-2} \psi_{jj'} ^ {(\bm{\gamma})}$ converges to a finite function with respect to $\bm{\gamma}$, as $n$ goes to infinity, uniformly for $\bm{\gamma}$ in the compact set $\Theta$, for all $j$ and $j'$;
		and $n ^ {-1} \operatorname{tr} \left\{\left[\partial ^ 2 \bm{\Omega} \left(\textbf{A}_{\bm{\Omega}}, \textbf{B}_{\bm{\Omega}}, \bm{\ell} \right) / \left(\partial \bm{\gamma}_j \partial \bm{\gamma}_{j'}\right) \right] \bm{\Sigma} \left(\textbf{A}_{\bm{\Sigma}}, \textbf{B}_{\bm{\Sigma}}, \bm{\ell} \right) \right\} ^ 2$ converges to $0$, as $n$ goes to infinity, uniformly for $\bm{\gamma}$ in the compact set $\Theta$, for all $j$ and $j'$.
		Given the consistency, $\widehat{\bm{\gamma}}$ is the unique ML estimator due to the result of Lemma 2 in \citet{Magnus1978}.
		
		Finally, the proof of (5) is straightforward.
	\end{proof}
	
	\begin{proof}[Simplifying the log-likelihood function]
		
		The vector version of the MAUD is given by $\textbf{y}_{(nR) \times 1} \sim N \left(\textbf{x}_{(nR) \times (Rp)} \bm{\beta}_{(Rp) \times 1}, \textbf{I}_n \otimes \bm{\Sigma}_{R \times R}\right)$, resulting in the log-likelihood function 
		\begin{equation*}
			\ell_n \left(\bm{\beta}, \bm{\gamma}; \textbf{x}, \textbf{y}\right)
			= - \frac{n R}{2} \log(2 \pi) - \frac{n}{2} \log \det (\bm{\Sigma})
			- \frac{n}{2} 
			\left[\frac{1}{n}\left(\textbf{y} - \textbf{x} \bm{\beta}\right) ^ \top \left(\textbf{I}_n \otimes \bm{\Sigma} ^ {-1}\right)
			\left(\textbf{y} - \textbf{x} \bm{\beta}\right)\right],
		\end{equation*}
		where we use the properties $\det \left(\textbf{A}_{n \times n} \otimes \textbf{B}_{m \times m} \right) = \left[\det \left(\textbf{A}\right)\right] ^ {m} \left[\det \left(\textbf{B}\right)\right] ^ {n}$ and $\left(\textbf{A} \otimes \textbf{B}\right) ^ {-1} = \textbf{A} ^ {-1} \otimes \textbf{B} ^ {-1}$.
		
		We observe that 
		
		\begin{equation*}
			\textbf{y}_{(nR) \times 1} - \textbf{x}_{(nR) \times (Rp)} \bm{\beta}_{(Rp) \times 1} = 
			\begin{pmatrix}
				\textbf{y}_1 - \left(\textbf{I}_{R} \otimes \textbf{x}_1 ^ \top \right) \bm{\beta} \\ \vdots \\ \textbf{y}_n - \left(\textbf{I}_{R} \otimes \textbf{x}_n ^ \top \right) \bm{\beta}
			\end{pmatrix}_{(nR) \times 1}
			= 
			\begin{pmatrix}
				\textbf{y}_1 - \mathfrak{B}_{R \times p} \textbf{x}_1 \\ \vdots \\ \textbf{y}_n - \mathfrak{B}_{R \times p} \textbf{x}_n
			\end{pmatrix}_{(nR) \times 1}.
		\end{equation*}
		It is because, for a fixed $i$,
		
		\begin{equation*}
			\begin{split}
				\textbf{y}_i - \left(\textbf{I}_{R} \otimes \textbf{x}_i ^ \top \right) \bm{\beta}
				& = \textbf{y}_i - 
				\begin{pmatrix}
					\textbf{x}_i ^ \top & & & \\
					& \textbf{x}_i ^ \top & & \\
					& & \ddots & \\
					& & & \textbf{x}_i ^ \top
				\end{pmatrix}_{R \times (Rp)}
				\bm{\beta} 
				= \textbf{y}_i - 
				\begin{pmatrix}
					\textbf{x}_i ^ \top \bm{\beta}_1 \\
					\textbf{x}_i ^ \top \bm{\beta}_2 \\
					\vdots \\
					\textbf{x}_i ^ \top \bm{\beta}_R
				\end{pmatrix}_{R \times 1}
				= \textbf{y}_i - 
				\begin{pmatrix}
					\bm{\beta}_1 ^ \top \textbf{x}_i \\
					\bm{\beta}_2 ^ \top \textbf{x}_i \\
					\vdots \\
					\bm{\beta}_R ^ \top \textbf{x}_i
				\end{pmatrix}_{R \times 1} \\
				& = \textbf{y}_i - \mathfrak{B}_{R \times p} \textbf{x}_i,
			\end{split}
		\end{equation*}
		where $\mathfrak{B} ^ \top$ is defined as $\left(\bm{\beta}_1, \ldots, \bm{\beta}_R \right) \in \mathbb{R} ^ {p \times R}$.
		Thus, 
		
		\begin{equation*}
			\begin{split}
				& \frac{1}{n}\left(\textbf{y} - \textbf{x} \bm{\beta}\right) ^ \top \left(\textbf{I}_n \otimes \bm{\Sigma} ^ {-1}\right)
				\left(\textbf{y} - \textbf{x} \bm{\beta}\right) \\
				& = \frac{1}{n}
				\begin{pmatrix}
					\left(\textbf{y}_1 - \mathfrak{B} \textbf{x}_1 \right) ^ \top, 
					\ldots, 
					\left(\textbf{y}_n - \mathfrak{B} \textbf{x}_n \right) ^ \top
				\end{pmatrix}_{1 \times (n R)}
				\begin{pmatrix}
					\bm{\Sigma} ^ {-1} & & & \\
					& \bm{\Sigma} ^ {-1} & & \\
					& & \ddots & \\
					& & & \bm{\Sigma} ^ {-1}
				\end{pmatrix}_{(nR) \times (nR)}
				\begin{pmatrix}
					\textbf{y}_1 - \mathfrak{B}\textbf{x}_1 \\
					\vdots \\ \textbf{y}_n - \mathfrak{B}\textbf{x}_n
				\end{pmatrix}_{(nR) \times 1} \\
				& = \frac{1}{n}
				\sum_{i = 1}^{n}
				\left(\textbf{y}_i - \mathfrak{B} \textbf{x}_i \right) ^ \top
				\bm{\Sigma} ^ {-1}
				\left(\textbf{y}_i - \mathfrak{B} \textbf{x}_i \right) \\
				& = \operatorname{tr}\left[
				\frac{1}{n}
				\sum_{i = 1}^{n}
				\left(\textbf{y}_i - \mathfrak{B} \textbf{x}_i \right)
				\left(\textbf{y}_i - \mathfrak{B} \textbf{x}_i \right) ^ \top
				\bm{\Sigma} ^ {-1}
				\right].
			\end{split}
		\end{equation*}
		Finally, we let $\textbf{S} = \textbf{S} \left(\bm{\beta}\right) = n ^ {-1} \sum_{i = 1}^{n}
		\left(\textbf{y}_i - \mathfrak{B} \textbf{x}_i \right)
		\left(\textbf{y}_i - \mathfrak{B} \textbf{x}_i \right) ^ \top \in \mathbb{R} ^ {R \times R}$ denote the residual matrix, 
		the log-likelihood function reduces to 
		
		\begin{equation*}
			\ell_n \left(\bm{\beta}, \bm{\gamma}; \textbf{x}, \textbf{y}\right)
			= - \frac{n R}{2} \log(2 \pi) + \frac{n}{2} \log \det (\bm{\Omega})
			- \frac{n}{2} 
			\operatorname{tr}\left(
			\textbf{S}
			\bm{\Omega}
			\right),
		\end{equation*}
		where $\bm{\Omega} = \bm{\Sigma} ^ {-1}$ denotes the precision matrix.
	\end{proof}
	
	\begin{proof}[Derivation of the score function]
		
		Plugging $\widehat{\bm{\beta}}$ into the log-likelihood function and letting $\textbf{S} = \textbf{S} \left(\widehat{\bm{\beta}}\right) = n ^ {-1} \sum_{i = 1}^{n}
		\left(\textbf{y}_i - \widehat{\mathfrak{B}} \textbf{x}_i \right)
		\left(\textbf{y}_i - \widehat{\mathfrak{B}} \textbf{x}_i \right) ^ \top \in \mathbb{R} ^ {R \times R}$ denote the residual matrix,		
		the log-likelihood function for $\bm{\gamma}$:
		\begin{equation*}
			\ell_n \left(\widehat{\bm{\beta}}, \bm{\gamma}; \textbf{x}, \textbf{y} \right)
			= \frac{n}{2}
			\left [
			- R \log(2 \pi)
			+ \log \det \left(\bm{\Omega}\right)
			- \operatorname{tr} \left(\textbf{S} \bm{\Omega}\right)
			\right ].
		\end{equation*}
		
		Before proceeding to the score function, we examine the following equality.
		Suppose $\textbf{M} \in \mathbb{R} ^ {R \times R}$ is a symmetric matrix and $\textbf{N}\left(\textbf{A}, \textbf{B}, \bm{\ell} \right)$ is a uniform-block matrix. 
		We partition $\textbf{M}$ by $\bm{\ell}$ into $\left(\textbf{M}_{gg'}\right)$, where $\textbf{M}_{gg'} \in \mathbb{R} ^ {L_g \times L_{g'}}$,
		
		\begin{equation*}
			\begin{split}
				\operatorname{tr}\left[\textbf{M} \textbf{N}(\textbf{A}, \textbf{B}, \bm{\ell})\right] 
				& = \operatorname{tr} \left[
				\begin{pmatrix}
					\textbf{M}_{11} & \textbf{M}_{12} & \dots & \textbf{M}_{1 G} \\
					\textbf{M}_{21} & \textbf{M}_{22} & \dots & \textbf{M}_{2 G} \\
					\vdots & \vdots & \ddots & \vdots \\
					\textbf{M}_{G1} & \textbf{M}_{G2} & \dots & \textbf{M}_{G G} 
				\end{pmatrix}
				\begin{pmatrix}
					\textbf{N}_{11} & \textbf{N}_{12} & \dots & \textbf{N}_{1 G} \\
					\textbf{N}_{21} & \textbf{N}_{22} & \dots & \textbf{N}_{2 G} \\
					\vdots & \vdots & \ddots & \vdots \\
					\textbf{N}_{G1} & \textbf{N}_{G2} & \dots & \textbf{N}_{G G} 
				\end{pmatrix}
				\right] \\
				& = \operatorname{tr}
				\begin{pmatrix}
					\sum_{g = 1}^{G} \textbf{M}_{1g} \textbf{N}_{g1} & 
					\sum_{g = 1}^{G} \textbf{M}_{1g} \textbf{N}_{g2} & 
					\dots & 
					\sum_{g = 1}^{G} \textbf{M}_{1g} \textbf{N}_{gG} \\
					%%%
					\sum_{g = 1}^{G} \textbf{M}_{2g} \textbf{N}_{g1} & 
					\sum_{g = 1}^{G} \textbf{M}_{2g} \textbf{N}_{g2} & 
					\dots & 
					\sum_{g = 1}^{G} \textbf{M}_{2g} \textbf{N}_{gG} \\
					\vdots & \vdots & \ddots & \vdots \\
					\sum_{g = 1}^{G} \textbf{M}_{Gg} \textbf{N}_{g1} & 
					\sum_{g = 1}^{G} \textbf{M}_{Gg} \textbf{N}_{g2} & 
					\dots & 
					\sum_{g = 1}^{G} \textbf{M}_{Gg} \textbf{N}_{gG} 
				\end{pmatrix} \\
				& = \operatorname{tr}
				\left(\sum_{g = 1}^{G} \sum_{g' = 1}^{G} \textbf{M}_{gg'} \textbf{N}_{g'g} \right) \\
				& = \sum_{g = 1}^{G} \sum_{g' = 1}^{G} \operatorname{tr}
				\left(\textbf{M}_{gg'} \textbf{N}_{g'g} \right).
			\end{split}
		\end{equation*}
		Using the definitions $\textbf{N}_{g'g} = a_{gg} \textbf{I}_{L_g} + b_{gg} \textbf{J}_{L_g}$ if $g' = g$ and $\textbf{N}_{g'g} = b_{g'g} \textbf{1}_{L_{g'} \times L_g}$ if $g' \neq g$, we have 
		
		\begin{equation*}
			\begin{split}
				& \operatorname{tr} \left(\textbf{M}_{gg'} \textbf{N}_{g'g} \right) \\
				& = 
				\begin{cases}
					\operatorname{tr} \left[\textbf{M}_{gg} \left(a_{gg} \textbf{I}_{L_g} + b_{gg} \textbf{J}_{L_g}\right) \right]
					= \operatorname{tr}\left[
					a_{gg} \textbf{M}_{gg} + b_{gg} \textbf{M}_{gg} \textbf{J}_{L_g}
					\right] = a_{gg} \operatorname{tr}\left(\textbf{M}_{gg}\right)
					+ b_{gg} \operatorname{sum} \left(\textbf{M}_{gg}\right), & g' = g \\
					\operatorname{tr} \left[\textbf{M}_{gg'} \left(b_{g'g} \textbf{1}_{L_{g'} \times L_g}\right)\right]
					= b_{g'g} \operatorname{sum} \left(\textbf{M}_{gg'}\right)
					= b_{gg'} \operatorname{sum} \left(\textbf{M}_{gg'}\right), & g' \neq g
				\end{cases}
			\end{split}
		\end{equation*}
		due to the following facts
		
		\begin{equation*}
			\begin{split}
				\operatorname{tr}\left(\textbf{M}_{gg} \textbf{J}_{L_g}\right)
				& = 
				\operatorname{tr}
				\begin{pmatrix}
					m_{1,1} ^ {(gg)} & m_{1,2} ^ {(gg)} & \dots & m_{1,L_g} ^ {(gg)} \\
					m_{2,1} ^ {(gg)} & m_{2,2} ^ {(gg)} & \dots & m_{2,L_g} ^ {(gg)} \\
					\vdots & \vdots & \ddots & \vdots \\
					m_{L_g,1} ^ {(gg)} & m_{L_g,2} ^ {(gg)} & \dots & m_{L_g,L_g} ^ {(gg)} 
				\end{pmatrix}_{L_g \times L_g}
				\begin{pmatrix}
					1 & 1 & \dots & 1 \\
					1 & 1 & \dots & 1 \\
					\vdots & \vdots & \ddots & \vdots \\
					1 & 1 & \dots & 1 
				\end{pmatrix}_{L_g \times L_g} \\
				& = \operatorname{tr}
				\begin{pmatrix}
					\sum_{j = 1}^{L_g} m_{1, j} ^ {(gg)} & \sum_{j = 1}^{L_g} m_{1, j} ^ {(gg)} & \dots & \sum_{j = 1}^{L_g} m_{1, j} ^ {(gg)} \\
					\sum_{j = 1}^{L_g} m_{2, j} ^ {(gg)} & \sum_{j = 1}^{L_g} m_{2, j} ^ {(gg)} & \dots & \sum_{j = 1}^{L_g} m_{2, j} ^ {(gg)} \\
					\vdots & \vdots & \ddots & \vdots \\
					\sum_{j = 1}^{L_g} m_{L_g, j} ^ {(gg)} & \sum_{j = 1}^{L_g} m_{L_g, j} ^ {(gg)} & \dots & \sum_{j = 1}^{L_g} m_{L_g, j} ^ {(gg)} 
				\end{pmatrix} \\
				& = \sum_{j = 1}^{L_g} \sum_{j' = 1}^{L_g} m_{j,j'} ^ {(gg)} \\
				& = \operatorname{sum} \left(\textbf{M}_{gg}\right); 
			\end{split}
		\end{equation*}
		and
		\begin{equation*}
			\begin{split}
				\operatorname{tr}\left(\textbf{M}_{gg'} \textbf{1}_{L_{g'} \times L_{g}}\right)
				& = \operatorname{tr}
				\begin{pmatrix}
					m_{1,1} ^ {(gg')} & m_{1,2} ^ {(gg')} & \dots & m_{1,L_{g'}} ^ {(gg')} \\
					m_{2,1} ^ {(gg')} & m_{2,2} ^ {(gg')} & \dots & m_{2,L_{g'}} ^ {(gg')} \\
					\vdots & \vdots & \ddots & \vdots \\
					m_{L_g,1} ^ {(gg')} & m_{L_g,2} ^ {(gg')} & \dots & m_{L_g,L_{g'}} ^ {(gg')}
				\end{pmatrix}_{L_g \times L_{g'}}
				\begin{pmatrix}
					1 & 1 & \dots & 1 \\
					1 & 1 & \dots & 1 \\
					\vdots & \vdots & \ddots & \vdots \\
					1 & 1 & \dots & 1 
				\end{pmatrix}_{L_{g'} \times L_g} \\
				& = 
				\operatorname{tr}
				\begin{pmatrix}
					\sum_{j = 1}^{L_{g'}} m_{1, j} ^ {(gg')} &
					\sum_{j = 1}^{L_{g'}} m_{1, j} ^ {(gg')} & 
					\dots & 
					\sum_{j = 1}^{L_{g'}} m_{1, j} ^ {(gg')} \\
					\sum_{j = 1}^{L_{g'}} m_{2, j} ^ {(gg')} &
					\sum_{j = 1}^{L_{g'}} m_{2, j} ^ {(gg')} & 
					\dots & 
					\sum_{j = 1}^{L_{g'}} m_{2, j} ^ {(gg')} \\
					\vdots & \vdots & \ddots & \vdots \\
					\sum_{j = 1}^{L_{g'}} m_{L_g, j} ^ {(gg')} &
					\sum_{j = 1}^{L_{g'}} m_{L_g, j} ^ {(gg')} & 
					\dots & 
					\sum_{j = 1}^{L_{g'}} m_{L_g, j} ^ {(gg')} \\
				\end{pmatrix} \\
				& = \sum_{j = 1}^{L_g} \sum_{j' = 1}^{L_{g'}} m_{j, j'} ^ {(gg')} \\
				& = \operatorname{sum} \left(\textbf{M}_{gg'}\right).
			\end{split}
		\end{equation*}
		
		Finally, we propose the following equality:
		
		\begin{equation*}
			\begin{split}
				\operatorname{tr}\left[\textbf{M}\textbf{N}(\textbf{A}, \textbf{B}, \bm{\ell})\right]
				& = \sum_{g = 1}^{G} \sum_{g' = 1}^{G} \operatorname{tr}
				\left(\textbf{M}_{gg'} \textbf{N}_{g'g} \right) \\
				& = \sum_{g = 1}^{G} \left[
				a_{gg} \operatorname{tr}\left(\textbf{M}_{gg}\right)
				+ b_{gg} \operatorname{sum} \left(\textbf{M}_{gg}\right)
				\right] + \sum_{g' \neq g} \left[
				b_{gg'} \operatorname{sum} \left(\textbf{M}_{gg'}\right)
				\right] \\
				& = \operatorname{sum} \left[
				\textbf{A} \odot \left(\operatorname{tr}\left(\textbf{M}_{gg}\right)\right)
				\right] + \operatorname{sum} \left[
				\textbf{B} \odot \left(\operatorname{sum}\left(\textbf{M}_{gg'}\right)\right)
				\right] \\
				& = \operatorname{sum}
				\begin{pmatrix}
					a_{11} \operatorname{tr}\left(\textbf{M}_{11}\right) & & & \\
					& a_{22} \operatorname{tr}\left(\textbf{M}_{22}\right) & & \\
					& & \ddots & \\
					& & & a_{GG} \operatorname{tr}\left(\textbf{M}_{GG}\right)
				\end{pmatrix} \\
				& + \operatorname{sum}
				\begin{pmatrix}
					b_{11} \operatorname{sum} \left(\textbf{M}_{11}\right) & 
					b_{12} \operatorname{sum} \left(\textbf{M}_{12}\right) & 
					\dots & 
					b_{1G} \operatorname{sum} \left(\textbf{M}_{1G}\right) \\
					b_{21} \operatorname{sum} \left(\textbf{M}_{21}\right) & 
					b_{22} \operatorname{sum} \left(\textbf{M}_{22}\right) & 
					\dots & 
					b_{2G} \operatorname{sum} \left(\textbf{M}_{2G}\right) \\
					\vdots & \vdots & \ddots & \vdots \\
					b_{G1} \operatorname{sum} \left(\textbf{M}_{G1}\right) & 
					b_{G2} \operatorname{sum} \left(\textbf{M}_{G2}\right) & 
					\dots & 
					b_{GG} \operatorname{sum} \left(\textbf{M}_{GG}\right) 
				\end{pmatrix} \\
				& = \operatorname{tr} \left[\textbf{A} \odot \operatorname{diag} \left(\operatorname{tr}\left(\textbf{M}_{11}\right), \ldots, \operatorname{tr}\left(\textbf{M}_{GG} \right)\right) \right]
				+ \operatorname{sum} \left[
				\textbf{B} \odot \left(\operatorname{sum} \left(\textbf{M}_{gg'}\right)\right)
				\right].
			\end{split}
		\end{equation*}
		
		Furthermore, if $\textbf{M} = \textbf{M}\left(\textbf{A}_{\textbf{M}}, \textbf{B}_{\bm{\textbf{M}}}, \bm{\ell} \right)$ is a uniform-block matrix, and $\textbf{M} = \textbf{N}\left(\textbf{A}_{\textbf{N}}, \textbf{B}_{\textbf{N}}, \bm{\ell} \right)$, then 
		
		\begin{equation*}
			\begin{split}
				& \operatorname{tr}\left[\textbf{M}\left(\textbf{A}_{\textbf{M}}, \textbf{B}_{\bm{\textbf{M}}}, \bm{\ell} \right)
				\textbf{N}\left(\textbf{A}_{\textbf{N}}, \textbf{B}_{\textbf{N}}, \bm{\ell} \right)\right] \\
				& = \operatorname{sum}
				\begin{pmatrix}
					a_{\textbf{N}, 11} \operatorname{tr}\left(\textbf{M}_{11}\right) & & & \\
					& a_{\textbf{N}, 22} \operatorname{tr}\left(\textbf{M}_{22}\right) & & \\
					& & \ddots & \\
					& & & a_{\textbf{N}, GG} \operatorname{tr}\left(\textbf{M}_{GG}\right)
				\end{pmatrix} \\
				& + \operatorname{sum}
				\begin{pmatrix}
					b_{\textbf{N}, 11} \operatorname{sum} \left(\textbf{M}_{11}\right) & 
					b_{\textbf{N}, 12} \operatorname{sum} \left(\textbf{M}_{12}\right) & 
					\dots & 
					b_{\textbf{N}, 1G} \operatorname{sum} \left(\textbf{M}_{1G}\right) \\
					b_{\textbf{N}, 21} \operatorname{sum} \left(\textbf{M}_{21}\right) & 
					b_{\textbf{N}, 22} \operatorname{sum} \left(\textbf{M}_{22}\right) & 
					\dots & 
					b_{\textbf{N}, 2G} \operatorname{sum} \left(\textbf{M}_{2G}\right) \\
					\vdots & \vdots & \ddots & \vdots \\
					b_{\textbf{N}, G1} \operatorname{sum} \left(\textbf{M}_{G1}\right) & 
					b_{\textbf{N}, G2} \operatorname{sum} \left(\textbf{M}_{G2}\right) & 
					\dots & 
					b_{\textbf{N}, GG} \operatorname{sum} \left(\textbf{M}_{GG}\right) 
				\end{pmatrix} \\
				& = \operatorname{sum}
				\begin{pmatrix}
					a_{\textbf{N}, 11} \left(a_{\textbf{M}, 11} + b_{\textbf{M}, 11}\right) L_1 & & & \\
					& a_{\textbf{N}, 22} \left(a_{\textbf{M}, 22} + b_{\textbf{M}, 22}\right) L_2 & & \\
					& & \ddots & \\
					& & & a_{\textbf{N}, GG} \left(a_{\textbf{M}, GG} + b_{\textbf{M}, GG}\right) L_G
				\end{pmatrix} \\
				& + \operatorname{sum}
				\begin{pmatrix}
					b_{\textbf{N}, 11} \left(L_1 a_{\textbf{M}, 11} + L_1 ^ 2 b_{\textbf{M}, 11} \right) & 
					b_{\textbf{N}, 12} b_{\textbf{M}, 12} L_1 L_2 & 
					\dots & 
					b_{\textbf{N}, 1G} b_{\textbf{M}, 1G} L_1 L_G \\
					b_{\textbf{N}, 21} b_{\textbf{M}, 21} L_2 L_1 & 
					b_{\textbf{N}, 22} \left(L_2 a_{\textbf{M}, 22} + L_2 ^ 2 b_{\textbf{M}, 22} \right) & 
					\dots & 
					b_{\textbf{N}, 2G} b_{\textbf{M}, 2G} L_2 L_G \\
					\vdots & \vdots & \ddots & \vdots \\
					b_{\textbf{N}, G1} b_{\textbf{M}, G1} L_G L_1 & 
					b_{\textbf{N}, G2} b_{\textbf{M}, G2} L_G L_2 & 
					\dots & 
					b_{\textbf{N}, GG} \left(L_G a_{\textbf{M}, GG} + L_G ^ 2 b_{\textbf{M}, GG} \right) 
				\end{pmatrix} \\
				& = \sum_{g = 1}^{G} a_{\textbf{N}, gg} \left(a_{\textbf{M}, gg} + b_{\textbf{M}, gg} \right) L_g 
				+ \sum_{g = 1}^{G} b_{\textbf{N}, gg} a_{\textbf{M}, gg} L_g
				+ \sum_{g = 1}^{G} \sum_{g' = 1}^{G}
				b_{\textbf{N}, gg'} b_{\textbf{M}, gg'} L_g L_{g'} \\
				& = \sum_{g = 1}^{G} a_{\textbf{N}, gg} a_{\textbf{M}, gg} L_g 
				+ \sum_{g = 1}^{G} a_{\textbf{N}, gg} b_{\textbf{M}, gg} L_g 
				+ \sum_{g = 1}^{G} b_{\textbf{N}, gg} a_{\textbf{M}, gg} L_g
				+ \sum_{g = 1}^{G} \sum_{g' = 1}^{G}
				b_{\textbf{N}, gg'} b_{\textbf{M}, gg'} L_g L_{g'} \\
				& = \operatorname{tr}\left(\textbf{A}_{\textbf{N}} \odot  \textbf{A}_{\textbf{M}} \odot \textbf{L} \right)
				+ \operatorname{tr} \left(\textbf{A}_{\textbf{N}} \odot \textbf{B}_{\textbf{M}} \odot \textbf{L}\right)
				+ \operatorname{tr} \left(\textbf{B}_{\textbf{N}} \odot \textbf{A}_{\textbf{M}} \odot \textbf{L} \right)
				+ \operatorname{sum} \left[\textbf{B}_{\textbf{N}} \odot \textbf{B}_{\textbf{M}} \odot \left(\bm{\ell} \bm{\ell} ^ \top \right) \right],
			\end{split}
		\end{equation*}
		where $\textbf{L} = \operatorname{diag}\left(L_1, \ldots, L_G \right)$,
		$\odot$ denotes the entry-wise Hadamard product,
		and $\bm{\ell} = \left(L_1, \ldots, L_G\right) ^ \top$ is the partition-size vector.
		
		Now, we derive the score function for $\bm{\gamma}$.
		Let $\bm{\gamma}_j \in \{\gamma_{11}, \ldots, \gamma_{1G}, \ldots, \gamma_{2G}, \ldots, \gamma_{GG}\}$ denote the $j$th component of $\bm{\gamma}$ for $j = 1, \ldots, G(G + 1) / 2$. 
		First, taking the first derivative of $\log \det (\bm{\Omega})$ with respect to $\bm{\gamma}_j$ can be written as 
		\begin{equation*}
			\begin{split}
				\frac{\partial \ell_n}{\partial \bm{\gamma}_j}
				& = \frac{n}{2} \bigg\{
				\operatorname{tr} \left[\bm{\Sigma} \left(\textbf{A}_{\bm{\Sigma}}, \textbf{B}_{\bm{\Sigma}}, \bm{\ell}\right) \frac{\partial \bm{\Omega}\left(\textbf{A}_{\bm{\Omega}}, \textbf{B}_{\bm{\Omega}}, \bm{\ell}\right)}{\partial \bm{\gamma}_j} \right] \\
				& - \frac{\partial }{\partial \bm{\gamma}_j}
				\left \{ \operatorname{tr} \left[\textbf{A}_{\bm{\Omega}} \odot \operatorname{diag} \left(\operatorname{tr}\left(\textbf{S}_{11}\right), \ldots, \operatorname{tr}\left(\textbf{S}_{GG} \right)\right) \right]
				+ \operatorname{sum} \left[
				\textbf{B}_{\bm{\Omega}} \odot \left(\operatorname{sum} \left(\textbf{S}_{gg'}\right)\right)
				\right] \right \} \bigg\} \\
				%%%
				& = \frac{n}{2} \bigg\{
				\operatorname{tr}\left[\left(\frac{\textbf{A}_{\bm{\Omega}}}{\partial \bm{\gamma}_j}\right) \odot \textbf{A}_{\bm{\Sigma}} \odot \textbf{L} \right]
				+ \operatorname{tr}\left[\left(\frac{\textbf{A}_{\bm{\Omega}}}{\partial \bm{\gamma}_j}\right) \odot \textbf{B}_{\bm{\Sigma}} \odot \textbf{L} \right]
				+ \operatorname{tr}\left[\left(\frac{\textbf{B}_{\bm{\Omega}}}{\partial \bm{\gamma}_j}\right) \odot \textbf{A}_{\bm{\Sigma}} \odot \textbf{L} \right] \\
				& + \operatorname{sum}\left[\left(\frac{\textbf{B}_{\bm{\Omega}}}{\partial \bm{\gamma}_j}\right) \odot \textbf{B}_{\bm{\Sigma}} \odot \left(\bm{\ell} \bm{\ell} ^ \top \right) \right] 
				- \operatorname{tr} \left[\left(\frac{\partial \textbf{A}_{\bm{\Omega}}}{\partial \bm{\gamma}_j}\right) \odot \operatorname{diag} \left(\operatorname{tr}\left(\textbf{S}_{11}\right), \ldots, \operatorname{tr}\left(\textbf{S}_{GG} \right)\right) \right] \\
				& - \operatorname{sum} \left[
				\left(\frac{\partial \textbf{B}_{\bm{\Omega}}}{\partial \bm{\gamma}_j}\right) \odot \left(\operatorname{sum} \left(\textbf{S}_{gg'}\right)\right)
				\right] \bigg\},
			\end{split}
		\end{equation*}
		where $\bm{\Omega} ^ {-1} \left(\textbf{A}_{\bm{\Omega}}, \textbf{B}_{\bm{\Omega}}, \bm{\ell}\right) = \bm{\Sigma} \left(\textbf{A}_{\bm{\Sigma}}, \textbf{B}_{\bm{\Sigma}}, \bm{\ell}\right)$ and 
		$\dfrac{\partial \bm{\Omega} \left(\textbf{A}_{\bm{\Omega}}, \textbf{B}_{\bm{\Omega}}, \bm{\ell}\right) }{\partial \bm{\gamma}_j}
		= \left(\dfrac{\partial \textbf{A}_{\bm{\Omega}}}{\partial \bm{\gamma}_j}\right) \circ \textbf{I}(\bm{\ell})
		+ \left(\dfrac{\partial \textbf{B}_{\bm{\Omega}}}{\partial \bm{\gamma}_j}\right) \circ \textbf{J}(\bm{\ell})$, with~\eqref{Eq:partial1} and~\eqref{Eq:partial2}. 	
	\end{proof}

	\section{THE CASE OF \texorpdfstring{$\bm{\Sigma}_{\bm{\epsilon}} \neq \textbf{I}_R$}{Sigma}}
	
	\label{Sec:case}
	
	We set a new parametric covariance matrix 
	\begin{equation*}
		\begin{split}
			\bm{\Sigma} = \left(\textbf{I}_R - \bm{\Upsilon}\right) ^ {-1} \bm{\Sigma}_{\epsilon} 
			\left(\textbf{I}_R - \bm{\Upsilon}\right) ^ {-1},
		\end{split}
	\end{equation*}
	where $\bm{\Upsilon} = \bm{\Upsilon} \left(\textbf{A}_{\bm{\Upsilon}}, \textbf{B}_{\bm{\Upsilon}}, \bm{\ell}\right)$ and $\bm{\Sigma}_{\bm{\epsilon}} = \bm{\Sigma}_{\bm{\epsilon}}\left(\textbf{A}_{\bm{\epsilon}}, \textbf{B}_{\bm{\epsilon}}, \bm{\ell} \right)$ with $\textbf{A}_{\bm{\Upsilon}} = \operatorname{diag}\left(- \gamma_{11}, \ldots, - \gamma_{GG}\right)$, $\textbf{B}_{\bm{\Upsilon}} = \left(\gamma_{gg'}\right)$, $\gamma_{g'g} = \gamma_{gg'}$, $\textbf{A}_{\bm{\epsilon}} = \operatorname{diag}\left(\omega_{11}, \ldots, \omega_{GG} \right)$, and $\textbf{B}_{\bm{\epsilon}} = \textbf{0}_{G \times G}$.
	
	Therefore, we can prove $\bm{\Sigma}$ and $\bm{\Omega}$ are also uniform-block matrices.
	\begin{equation*}
		\begin{split}
			\bm{\Omega} \left(\textbf{A}_{\bm{\Omega}}, \textbf{B}_{\bm{\Omega}}, \bm{\ell} \right) & = \left(\textbf{I}_R - \bm{\Upsilon}\right) \bm{\Sigma}_{\epsilon} ^ {-1}
			\left(\textbf{I}_R - \bm{\Upsilon}\right); \\
			%%%
			%\textbf{A}_1 & = \textbf{I}_{G} - \textbf{A}_{\bm{\Upsilon}}, 
			%\textbf{B}_1 = - \textbf{B}_{\bm{\Upsilon}}, \quad \left(\textbf{I}_R - \bm{\Upsilon}\right) \\
			%%%
			%\textbf{A}_2 & = \textbf{A}_{\bm{\epsilon}} ^ {-1}, 
			%\textbf{B}_2 = \textbf{0}_{G \times G}, \quad
			%\bm{\Sigma}_{\epsilon} ^ {-1} \\
			%%%
			%\textbf{A}_3 & = \textbf{A}_1 \textbf{A}_2, 
			%\textbf{B}_3 = \textbf{A}_1 \textbf{B}_2 + \textbf{B}_1 \textbf{A}_2 + \textbf{B}_1 \textbf{R} \textbf{B}_2, \quad
			%\left(\textbf{I}_R - \bm{\Upsilon}\right) \bm{\Sigma}_{\epsilon} ^ {-1} \\
			%%%
			\textbf{A}_{\bm{\Omega}} 
			%& = \textbf{A}_3 \textbf{A}_1
			%= \textbf{A}_1 \textbf{A}_2 \textbf{A}_1
			%= \left(\textbf{I}_{G} - \textbf{A}_{\bm{\Upsilon}}\right)
			%\textbf{A}_{\bm{\epsilon}} ^ {-1}
			%\left(\textbf{I}_{G} - \textbf{A}_{\bm{\Upsilon}}\right)
			%= \left(\textbf{A}_{\bm{\epsilon}} ^ {-1} - \textbf{A}_{\bm{\Upsilon}} \textbf{A}_{\bm{\epsilon}} ^ {-1}\right)
			%\left(\textbf{I}_{G} - \textbf{A}_{\bm{\Upsilon}}\right) \\
			%& = \textbf{A}_{\bm{\epsilon}} ^ {-1} - \textbf{A}_{\bm{\epsilon}} ^ {-1} \textbf{A}_{\bm{\Upsilon}}
			%- \textbf{A}_{\bm{\Upsilon}} \textbf{A}_{\bm{\epsilon}} ^ {-1}
			%+ \textbf{A}_{\bm{\Upsilon}} \textbf{A}_{\bm{\epsilon}} ^ {-1} \textbf{A}_{\bm{\Upsilon}} \\
			& = \textbf{A}_{\bm{\epsilon}} ^ {-1} 
			- 2\textbf{A}_{\bm{\Upsilon}} \textbf{A}_{\bm{\epsilon}} ^ {-1}
			+ \textbf{A}_{\bm{\Upsilon}} ^ 2 \textbf{A}_{\bm{\epsilon}} ^ {-1}, \\
			%%%
			\textbf{B}_{\bm{\Omega}} 
			%& = \textbf{A}_3 \textbf{B}_1 + \textbf{B}_3 \textbf{A}_1 + \textbf{B}_3 \textbf{R} \textbf{B}_1 \\
			%& = (\textbf{A}_1 \textbf{A}_2) 
			%(- \textbf{B}_{\bm{\Upsilon}}) + 
			%(\textbf{A}_1 \textbf{B}_2 + \textbf{B}_1 \textbf{A}_2 + \textbf{B}_1 \textbf{R} \textbf{B}_2) 
			%\textbf{A}_1 + 
			%(\textbf{A}_1 \textbf{B}_2 + \textbf{B}_1 \textbf{A}_2 + \textbf{B}_1 \textbf{R} \textbf{B}_2) 
			%\textbf{R} 
			%(- \textbf{B}_{\bm{\Upsilon}}) \\
			%%%%
			%& = (\textbf{A}_1 \textbf{A}_2) 
			%(- \textbf{B}_{\bm{\Upsilon}}) + 
			%(\textbf{A}_1 \textbf{0}_{G \times G} + \textbf{B}_1 \textbf{A}_2 + \textbf{B}_1 \textbf{R} \textbf{0}_{G \times G}) 
			%\textbf{A}_1 + 
			%(\textbf{A}_1 \textbf{0}_{G \times G} + \textbf{B}_1 \textbf{A}_2 + \textbf{B}_1 \textbf{R} \textbf{0}_{G \times G}) 
			%\textbf{R} 
			%(- \textbf{B}_{\bm{\Upsilon}}) \\
			%%%%
			%& = - \textbf{A}_1 \textbf{A}_2
			%\textbf{B}_{\bm{\Upsilon}} + 
			%\textbf{B}_1 \textbf{A}_2
			%\textbf{A}_1 
			%- \textbf{B}_1 \textbf{A}_2
			%\textbf{R} \textbf{B}_{\bm{\Upsilon}} \\
			%%%%
			%& = - (\textbf{I}_{G} - \textbf{A}_{\bm{\Upsilon}}) \textbf{A}_{\bm{\epsilon}} ^ {-1}
			%\textbf{B}_{\bm{\Upsilon}} + 
			%(- \textbf{B}_{\bm{\Upsilon}}) \textbf{A}_{\bm{\epsilon}} ^ {-1}
			%(\textbf{I}_{G} - \textbf{A}_{\bm{\Upsilon}})
			%- (- \textbf{B}_{\bm{\Upsilon}}) \textbf{A}_{\bm{\epsilon}} ^ {-1}
			%\textbf{R} \textbf{B}_{\bm{\Upsilon}} \\
			%%%
			& = 
			-\textbf{A}_{\bm{\epsilon}} ^ {-1} \textbf{B}_{\bm{\Upsilon}}
			+ \textbf{A}_{\bm{\Upsilon}} \textbf{A}_{\bm{\epsilon}} ^ {-1} \textbf{B}_{\bm{\Upsilon}}
			- \textbf{B}_{\bm{\Upsilon}} \textbf{A}_{\bm{\epsilon}} ^ {-1}
			+ \textbf{B}_{\bm{\Upsilon}} \textbf{A}_{\bm{\epsilon}} ^ {-1} \textbf{A}_{\bm{\Upsilon}}
			+ \textbf{B}_{\bm{\Upsilon}} \textbf{A}_{\bm{\epsilon}} ^ {-1}
			\textbf{L} \textbf{B}_{\bm{\Upsilon}}.
		\end{split}
	\end{equation*}
	We can observe that both $\textbf{A}_{\bm{\Omega}}$ and $\textbf{B}_{\bm{\Omega}}$ are symmetric.

	\newpage
	
	\bibliographystyle{biom}	
	
	\bibliography{ShareReferences}